\documentclass{emulateapj}

\usepackage{amsmath,natbib, float, color}
\newcommand*\mean[1]{\bar{#1}}

\newcommand*\Msolarh[0]{\mathrm{M_{\odot}}\, h^{-1}}
\newcommand*\Mpch[0]{\mathrm{Mpc} \, h^{-1}}
\newcommand*\Gpch[0]{\mathrm{Gpc} \, h^{-1}}

\newcommand*\PLM[0]{$\mathrm{PL_\mathrm{P}}$} 		
\newcommand*\PLI[0]{$\mathrm{PL_\mathrm{C}}$}	
\newcommand*\MLv[0]{$\mathrm{ML_v}$}  	
\newcommand*\MLvP[0]{$\mathrm{ML_{v,\,P}}$}  			
\newcommand*\MLR[0]{$\mathrm{ML_R}$}				
\newcommand*\MLvR[0]{$\mathrm{ML_{v,R}}$}			
\newcommand*\MLvsR[0]{$\mathrm{ML_{v,\sigma,R}}$}		
\newcommand*\memb[0]{Pure}	
\newcommand*\interloper[0]{Contaminated}

\newcommand*\vlos[0]{v_{\mathrm{los}}}
\newcommand*\vsig[0]{|\vlos|/\sigma_v}
\newcommand*\kms[0]{\mathrm{km \, s^{-1}}}
\newcommand*\scut[0]{$s_{\mathrm{cut}}$}
\newcommand*\vcut[0]{$v_{\mathrm{cut}}$}

\newcommand{\possessivecite}[1]{\citeauthor{#1}'s \citeyear{#1}}

\shorttitle{ML Cluster Mass}
\shortauthors{Ntampaka et al.}

\begin{document}

\title{Dynamical Mass Measurements of Contaminated Galaxy Clusters \\ Using Machine Learning}

\author{M. Ntampaka\altaffilmark{1}, H. Trac\altaffilmark{1}, D.J. Sutherland\altaffilmark{2}, S. Fromenteau\altaffilmark{1}, B. P\'{o}czos\altaffilmark{2}, J. Schneider\altaffilmark{2}}
\email{ntampaka@cmu.edu}
\altaffiltext{1}{McWilliams Center for Cosmology, Department of Physics, Carnegie Mellon University, Pittsburgh, PA 15213}
\altaffiltext{2}{School of Computer Science, Carnegie Mellon University, Pittsburgh, PA 15213}
\affil{}

\begin{abstract}
We study dynamical mass measurements of galaxy clusters contaminated by interlopers and show that a modern machine learning (ML) algorithm can predict masses by better than a factor of two compared to a standard scaling relation approach. We create two mock catalogs from Multidark's publicly available $N$-body MDPL1 simulation, one with perfect galaxy cluster membership information and the other where a simple cylindrical cut around the cluster center allows interlopers to contaminate the clusters.  In the standard approach, we use a power-law scaling relation to infer cluster mass from galaxy line-of-sight (LOS) velocity dispersion. Assuming perfect membership knowledge, this unrealistic case produces a wide fractional mass error distribution, with a width of $\Delta\epsilon\approx0.87$. Interlopers introduce additional scatter, significantly widening the error distribution further ($\Delta\epsilon\approx2.13$). We employ the support distribution machine (SDM) class of algorithms to learn from distributions of data to predict single values.  Applied to distributions of galaxy observables such as LOS velocity and projected distance from the cluster center, SDM yields better than a factor-of-two improvement ($\Delta\epsilon\approx0.67$) {for the contaminated case}. Remarkably, SDM applied to contaminated clusters is better able to recover masses than even the scaling relation approach applied to uncontaminated clusters. We show that the SDM method more accurately reproduces the cluster mass function, making it a valuable tool for employing cluster observations to evaluate cosmological models. \\ 

\end{abstract}


\section{Introduction}
\label{sec:intro}

{Galaxy clusters are the most massive gravitationally-bound systems in the Universe.  They are dark matter dominated, and have halos of mass $\gtrsim 10^{14}\ \Msolarh$.  {The majority of m}ultiple-wavelength observations do not directly probe the dark matter distribution, but the baryonic component of clusters: the hot gas and tens to thousands of galaxies contained within the halo.  Clusters have complex substructure and internal dynamics, and grow through hierarchical merging and the accretion of matter from the cosmic web.  Cluster abundance as a function of mass and redshift is sensitive to the underlying dark matter and dark energy content of the Universe and can be used to test cosmological models. See \cite{2005RvMP...77..207V} and \cite{2011ARA&A..49..409A} for a review.}

{While measurements of cluster masses can be employed to constrain cosmological parameters 
\citep[e.g.][]{2003A&A...398..867S, 2009ApJ...691.1307H, 2009ApJ...692.1060V, 2010ApJ...708..645R,2010MNRAS.406.1773M, 2010ApJ...722.1180V, 2011ApJ...732...44S, 2011ARA&A..49..409A, 2014A&A...571A..20P, 2015MNRAS.446.2205M}, capitalizing on clusters as cosmological probes requires a large, well-defined sample of cluster observations, a connection linking the observations of the baryonic component to the underlying dark matter, and a good understanding of the intrinsic scatter in the mass-observable relationship.  A variety of methods connecting observables to cluster mass exist, utilizing observations across multiple wavelengths.} A subset of these techniques, broadly labeled dynamical mass measurements, are based on measurements of galaxy kinematics.  Dynamical mass measurements utilize line-of-sight (LOS) velocities of the galaxies within the virial radius of the cluster, and may also take advantage of the unvirialized matter falling toward the cluster.   

The virial theorem approach considers cluster members' LOS velocity dispersion, $\sigma_v$.  This method scales halo mass, $M$, with $\sigma_v$ as a power law and famously led to \possessivecite{zwicky1933rotverschiebung} discovery of dark matter in the Coma cluster.  Dynamical mass measurements based on the virial theorem continue to be used to determine cluster masses \citep[e.g.][]{2010ApJ...721...90B, 2010ApJ...715L.180R, 2013ApJ...772...25S, 2014ApJ...792...45R, 2015ApJ...799..214B}.  {\cite{2014MNRAS.441.1513O} and \cite{2015MNRAS.449.1897O} provide a comparison of several dynamical mass techniques based on galaxy observables.}  
{Even when cluster membership is perfectly and fully known, there is s}catter in the $M(\sigma_v)$ scaling relation{.  This} can be attributed to both physical {effects and selection} effects, including halo environment and triaxiality \citep[e.g.][]{White:2010ab, 2013ApJ...772...47S,  2013A&A...559A..89W, 2014arXiv1405.0284S}, projection effects \citep[e.g.][]{2012MNRAS.419.1017C, 2012MNRAS.426.1829N}, mass-dependent tidal disruption \citep[e.g.][]{2013MNRAS.430.2638M}, the degree of relaxedness of the cluster \citep[e.g.][]{Evrard:2008aa, 2011MNRAS.413L..81R}, and galaxy selection strategy 
{\citep[e.g.][]{2013MNRAS.434.2606O, 2013ApJ...772...47S, 2013MNRAS.436..460W}}.
  Halos undergoing mergers or matter accretion possess a telltale wide, flat velocity probability distribution function (PDF) \citep{2011MNRAS.413L..81R}.  {Impure, incomplete cluster membership catalogs increase scatter in the $M(\sigma_v)$ relationship further.  Reducing errors in cluster mass measurements is essential for applying clusters as cosmological probes.}

The galaxy dynamics beyond the virial radius of the cluster is likewise informative, and nearby, unvirialized matter can also be used for cluster mass measurements.  
The caustic technique employs infalling matter and galaxy velocities to determine a mass profile 
{\citep[e.g.][]{2003ApJ...585..205B, 2011MNRAS.412..800S, 2013ApJ...768L..32G}}
 and can be applied to determine cluster masses \citep[e.g.][]{2006AJ....132.1275R, 2013ApJ...767...15R, 2013ApJ...764...58G}, performing well even in the case of merging halos \citep[e.g.][]{2003AJ....126.2152R}.  
Further, the nonvirialized infalling matter beyond the virial radius provides cues which can be used to infer a cluster's mass \citep[e.g.][]{2013MNRAS.431.3319Z, 2014MNRAS.442.1887F}.

{A machine learning approach to dynamical mass measurements was explored in \cite{Ntampaka2015}.  Here, we built on the virial theorem's simple $M(\sigma_v)$ power law to take advantage of the entire LOS velocity PDF for mock observations with pure and complete cluster membership information, using all relevant substructure within the $R_\mathrm{200c}$ of each cluster.  Taking full advantage of the velocity PDF was achieved by applying a nonparametric machine learning (ML) approach to a PDF of LOS velocities from a mock cluster catalog.  By employing support distribution machines (SDMs), an ML class of algorithms that learns from a distribution to predict a scalar, the full velocity PDF was used to improve mass predictions.  A traditional power-law scaling relation yielded a wide fractional mass error distribution (see equation \ref{eq:fracerror}) and extended high-error tails.  SDMs trained on LOS velocities resulted in almost a factor-of-two reduction in mass errors compared to the traditional approach, substantially reducing the number of severely over- and underestimated halo masses in the ideal case with pure and complete cluster membership information. }

{ However, the idealized catalog used in this case did not account for a primary source of error in dynamical mass measurements: interloper galaxies in the fore- or background of the true cluster, appearing to be cluster members.  In an ideal cluster catalog, all cluster members are known (complete) and the observations contain only true members (pure).  Cluster observations that are impure due to contamination by interlopers are subject to additional scatter in the $M(\sigma_v)$ relationship \citep[e.g.][]{2010A&A...520A..30M}, and a variety of methods have been developed to remove interloper galaxies from the sample \citep[e.g.][]{1996ApJ...473..670F, 2007MNRAS.379..867V, 2013MNRAS.429.3079M, 2015MNRAS.449.3082P}}

In this follow-up paper, we explore how a more realistically-prepared mock catalog influences both the $M(\sigma_v)$ scaling relation as well as the SDM predictions of cluster mass. Cluster members are selected within a cylinder defined by a projected radius in the plane of the sky and a radial velocity along the line-of-sight.  This technique produces a catalog of spectroscopic member catalogs that are impure, containing interloping galaxies that appear to be cluster members but do not reside within the virial radius of the cluster.  They are also incomplete, excluding some true cluster members from the sample.  
 
In Sec.~\ref{sec:methods}, we discuss our methods:  the simulation (\ref{sec:sim}), mock observation (\ref{sec:mock}), power-law scaling relation (\ref{sec:PL}), and SDM implementation (\ref{sec:SDM}).  Results are presented in Sec.~\ref{sec:results} and discussed in Sec.~\ref{sec:discussion}.  We present a summary of our findings in Sec.~\ref{sec:conclusions}.  Finally, we explore how changes to our mock catalog affect power law and ML results in the Appendix (Sec.~\ref{sec:appendix}).

\section{Methods}
\label{sec:methods}

\subsection{Simulation}
\label{sec:sim}

The mock cluster catalog is created from the publicly available Multidark MDPL1 simulation\footnote{http://www.cosmosim.org/}.  Multidark is an $N$-body simulation containing $3840^3$ particles in a box of length $1\ h^{-1}\rm{Gpc}$ and a mass resolution of  $1.51\times10^9\, \Msolarh$.  Multidark was run using the L-Gadget2 code.  It utilizes a $\Lambda$CDM cosmology, with cosmological parameters consistent with Planck data \citep{2014A&A...571A..16P}:  $\Omega_{\Lambda} = 0.69$, $\Omega_m = 0.31$, $\Omega_b = 0.048$, $h = 0.68$, ${n=0.96}$, and $\sigma_8 = 0.82$.

Halos are identified by Multidark's BDMW algorithm, which uses a bound density maximum (BDM) spherical overdensity halo finder with halo average density equal to 200 times the critical density of the Universe, denoted $M$.  All halos and subhalos at redshift $z=0$ with mass  $M\geq10^{12} \,\Msolarh$ are included in our sample.  For more information on the Multidark simulation and BDMW halo finder, see \cite{1997astro.ph.12217K, 2013AN....334..691R, 2014arXiv1411.4001K} and references therein.

\subsection{Mock Observations}
\label{sec:mock}

\begin{deluxetable*}{llcccccccc}
\tabletypesize{\scriptsize}

\tablecaption{{Catalog Summary} \label{table:catalog}}
\tablewidth{0pt}
\tablehead{
\colhead{Catalog Name} & \colhead{Type} & \colhead{Min. Halo Mass} & \colhead{$R_\mathrm{aperture}$} & \colhead{\vcut} &
\colhead{\scut} & \colhead{Projections per} & \colhead{Total} &
\colhead{$\sigma_{15}$} & \colhead{$\alpha$} \\
\colhead{} & \colhead{} & \colhead{ $(\Msolarh)$} & \colhead{$(\Mpch)$} & \colhead{$(\kms)$} &
\colhead{} & \colhead{Unique Halo} & \colhead{Projections} &
\colhead{$(\kms)$} & \colhead{} 
}

\startdata
\memb{} 						& Train		& $1\times 10^{14} $ 			& -								&-							&-						& varies 			& 15000 							& 1244 							& 0.382 	\\[1.5ex]
\memb{}						& Test 		& $3\times 10^{14} $ 			& - 								&-							&-						& 3 								& 6834							&-  								&-  		\\[1.5ex]
\memb{}				& High Mass Test 		& $7\times 10^{14} $ 			& - 								&-							&-						& 3 								& 945							&-  								&-  		\\[1.5ex]

\interloper{}					& ML Train 	& $1\times 10^{14} $ 			& 1.6 							&2500						&2.0						& varies 			& 15000 							& -							& -	\\[1.5ex]
\interloper{}					& PL Train 	& $3\times 10^{14} $ 			& 1.6 							&2500						&2.0						& varies 			& 10213							& 753							& 0.359	\\[1.5ex]

\interloper{}					& Test 		& $3\times 10^{14} $				& 1.6 							&2500						&2.0						& 3 								& 7449 							&-  								&- 		\\[1.5ex]
\interloper{}					& High Mass Test 	& $7\times 10^{14} $			& 1.6 							&2500						&2.0						& 3 								& 951 							&-  								&- 		\\[1.5ex]

\enddata
\tablecomments{For the \memb{} Catalogs, cluster radius and member galaxies are known.  For further details on the creation of this catalog, see \cite{Ntampaka2015}.}
\end{deluxetable*}

{Two mock observations are created:   \memb{} and \interloper{}.  For each of these two mock observations, a train sample and a test sample are made.} The \memb{} Catalog is ideal, in that all cluster members above $M_\mathrm{sub}=10^{12}\,\Msolarh$ within $R_\mathrm{200}$ are included in the catalog.  The train catalog has a flat mass function, with 5028 unique halos with $M\geq10^{14}\,\Msolarh$.  Halos in this catalog contributes multiple lines of sight each such that low- and high mass clusters are represented in equal measures.  The test catalog has 2278 unique halos with a lower mass cut of $M\geq3\times10^{14}\,\Msolarh$, and each unique halo contributes exactly three lines of sight each. It is discussed in further detail in \cite{Ntampaka2015}.

{In contrast with the \memb{} Catalog, the  \interloper{} Catalog includes more realistic observational selection effects.  It employs a simple, cylindrical cut around each cluster, allowing interlopers to contaminate the sample.  As with the \memb{} Catalog, the  \interloper{} Catalog has both a train catalog with a flat mass function, as well as a test catalog that uses three lines of sight per cluster.}

{The \interloper{} Catalog is constructed in the following way:  each halo and subhalo is assumed to represent an observable galaxy, with the galaxy inheriting its host's position and velocity.  A simple cut is made around each cluster, allowing for interlopers to contaminate the cluster observation.}  To allow for interlopers across the box edge, the entire simulation box is padded with a $200 \, \Mpch$-thick slice from across the periodic boundary to make a cube with length $1.4\, \Gpch$.  This cubic mock observation will be used to create a mock cluster catalog that incorporates known observational selection effects.  

An intentionally-simplistic cylindrical cut is made around each cluster center.  Only halos with $M\geq10^{14}\, \Msolarh$ with centers that reside within the original $1\, \Gpch$box volume are considered to be ``cluster candidates.''  Following \cite{2014MNRAS.441.1513O}, true cluster centers are assumed to be known by the observer.   Following \cite{2007A&A...466..437W}, the observer is placed $100 \, \mathrm{Mpc}$ from the center of the cluster along the chosen line-of-sight.   

The full 3D galaxy velocity and position information is reduced, then, to what can be observed along this line-of-sight: plane-of-sky $x'$- and $y'$-positions and LOS velocities.  A galaxy's net velocity, $v$, is given by the sum of the peculiar velocity plus the Hubble flow.  An initial cylindrical cut defined by a circular aperture with radius $R_\mathrm{aperture}$ about the cluster center in the plane of the sky and a LOS initial velocity cut of $v_\mathrm{cut}$ about the expected hubble flow velocity of an object located at a distance of $100 \, \mathrm{Mpc}$ from the observer.  

The cylinder $R_\mathrm{aperture}$ and $v_\mathrm{cut}$ values are chosen to correspond with the radius and $2\sigma_v$, respectively, of a $1\times10^{15}\, \Msolarh$ cluster.  The radius of a cluster of this mass is $1.6 \, \Mpch$.  The $2\sigma_v$ is informed by the best fit power law found in \cite{Ntampaka2015}, giving twice a typical velocity dispersion of true cluster members of $2\sigma_v \approx 2500 \, \kms$ for a cluster of mass $1\times10^{15}\, \Msolarh$.  These parameters are noted in Table \ref{table:catalog}.  A more thorough exploration of how $R_\mathrm{aperture}$ and $v_\mathrm{cut}$ choices affect cluster mass predictions is presented in the Appendix (Sec.~\ref{sec:appendix}).

This initial cylinder is pared iteratively in velocity space, with outliers beyond $2 \sigma_v$ of the mean velocity being omitted from the sample.  Here, $\sigma_v$ denotes the standard deviation of all LOS velocities of the galaxies that reside in the cylinder.  This paring occurs until convergence is reached or until fewer than 20 members remain.  Clusters with at least 20 members remaining are added to the cluster catalog.

In order to create a representative training sample of how the rare, high-mass clusters might appear when viewed from any direction, the entire box is rotated and this process is repeated.  The first three rotations are chosen so that the observer views along the box $x$-~, \mbox{$y$-,} and $z$-directions.  The remaining rotations are chosen randomly on the surface of the unit sphere.  To create the \interloper{} Train Catalog, $1000$ such rotations are performed.

The Train Catalog includes halos with $M\geq1\times10^{14}\, \Msolarh$.  It is created with a flat mass function, such that there are exactly $1000$ training clusters in each $0.1 dex$ mass bin.  In bins with fewer than $1000$ clusters, this is done by assembling many LOS views of rare halos.  In mass bins with more than $1000$ clusters, clusters are rank ordered by mass and evenly removed from the training sample.

\begin{figure*}[!t]
  \centering
  \includegraphics[width=\textwidth]{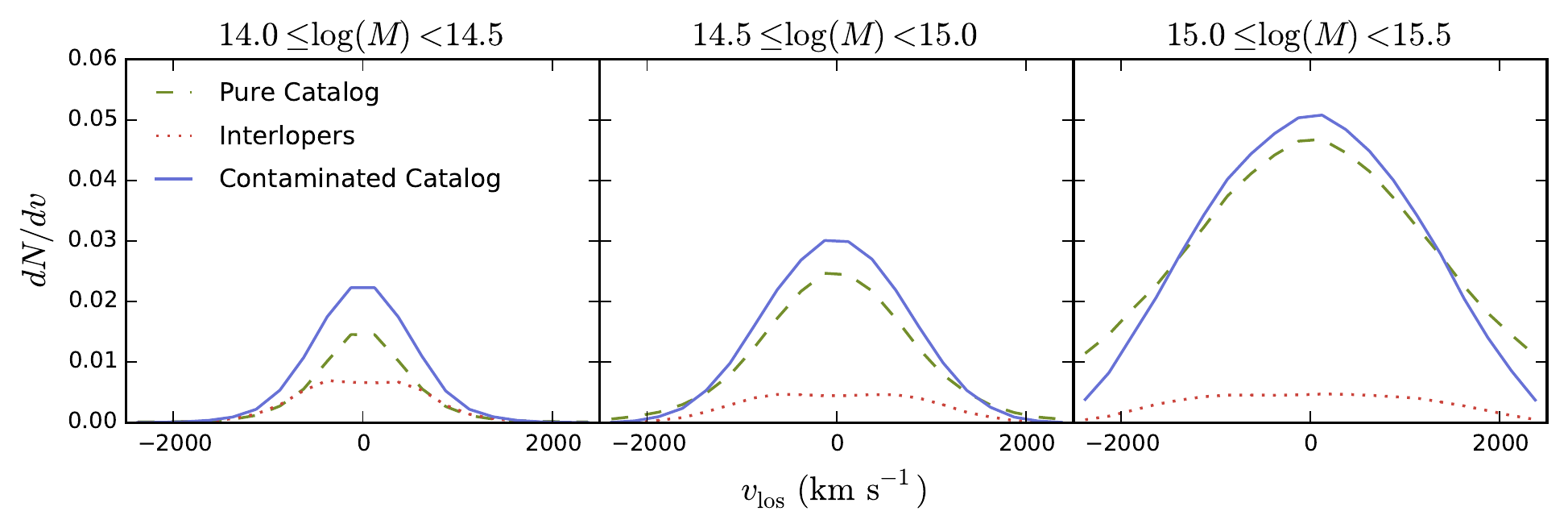}
  \includegraphics[width=\textwidth]{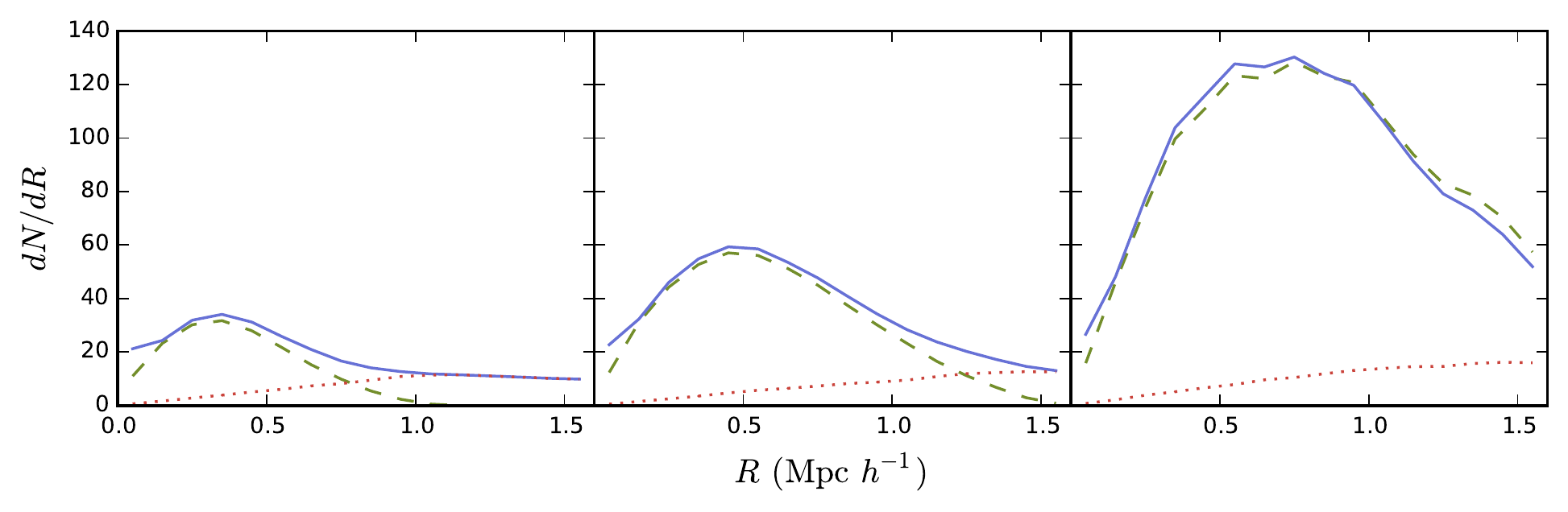}
  \caption{Top:  Average distribution of galaxy LOS velocities from stacked clusters in three $\log[M \, (\Msolarh)]$ bins, in increasing mass from left to right.  While the \memb{} Catalog  (green dashed) consists solely of galaxies residing within the virial radius of the cluster, the \interloper{} Catalog (blue solid) contains contaminating interlopers (red dotted) and excludes some true cluster members.  In the top right panel, the exclusion of true cluster members is evident where the blue solid line dips below green dashed.  
  Bottom:  Average distribution of galaxy projected radii from the cluster center.  Both $\vlos$ and $R$ distributions change shape and amplitude with cluster mass, even for the \interloper{} Catalog; this mass-dependent shape can be exploited by a distribution-to-scalar ML technique to learn cluster masses from distributions of data like the examples shown here. }
  \label{fig:intcut}
\end{figure*}

In contrast with the \interloper{} Train Catalog, the \interloper{} Test Catalog contains exactly three LOS views of every halo: the box $x$-~, $y$-~, and $z$-directions.  Because boundary effects are expected near the edge of the training sample, a minimum mass cut of $M\geq3\times10^{14}\, \Msolarh$ is applied to the test catalogs.  The single most massive halo has a mass that will necessarily lie outside of the training sample, and therefore is omitted from the test catalogs as well.  

In summary, the {\interloper{} Catalog} is created in the following manner:

\begin{enumerate}
	\item All halos and subhalos with mass greater than $10^{12} \, \Msolarh$ {are assumed to represent a galaxy, with the galaxy inheriting its host's position and velocity.}
	\item Halos with mass greater than $10^{14}\, \Msolarh$ are considered ``cluster candidates.'' 
	\item \label{startpare}A cluster candidate's center is assumed to be known, and an observer is placed $100\, \mathrm{Mpc}$ from the cluster.  
	\item All galaxies in the box are given an appropriate velocity that includes both Hubble flow and peculiar velocities.
	\item A cylinder is cut around the cluster candidate center; this cylinder is defined by an aperture radius, $R_\mathrm{aperture}$, and a LOS velocity cut, $v_\mathrm{cut}$.
	\item Galaxies outside of mean galaxy velocity $\pm \, 2\sigma_v$ are iteratively removed from this cylinder until convergence is reached.
	\item \label{endpare} This is repeated for all massive halos in the box, and those with at least 20 members remaining are kept in the sample.
	\item The box is rotated, and steps \ref{startpare}-\ref{endpare} are repeated.
	\item The \interloper{} Train Catalog is made of multiple LOS projections, up to 1000 for the highest-mass cluster.  The number of projections per unique halo is chosen to create a flat mass function for the Train Catalog.
	\item The \interloper{} Test Catalog is made of the first three ($x$-~, $y$-~, and $z$-directions) views of all halos above $M=3\times10^{14}\, \Msolarh$.  The most massive halo is also excluded from the Test Catalog.  
\end{enumerate}


Figure \ref{fig:intcut} shows the average $\vlos$ and $R$ distributions for the Train Catalogs, divided into three $\log[M(\Msolarh)]$ bins.  The \memb{} Catalog is pure, in that there are no interlopers contaminating the galaxy clusters.  It is also complete, in that all galaxies within the cluster $R_{200}$ are known.   In contrast,  the \interloper{} Catalog includes interlopers and excludes some true cluster members.  The shape of $\vlos$ and $R$ distributions are mass-dependent, and this dependence on cluster mass can be utilized in mass predictions.  In Sec.~\ref{sec:SDM}, we will explore ways to predict cluster mass by exploiting these mass-dependent distributions using a distribution-to-scalar machine learning technique.
\\

 \subsection{Power Law}
 \label{sec:PL}
 
 \begin{figure}[!tb]
  \centering
  \includegraphics[width=0.5\textwidth]{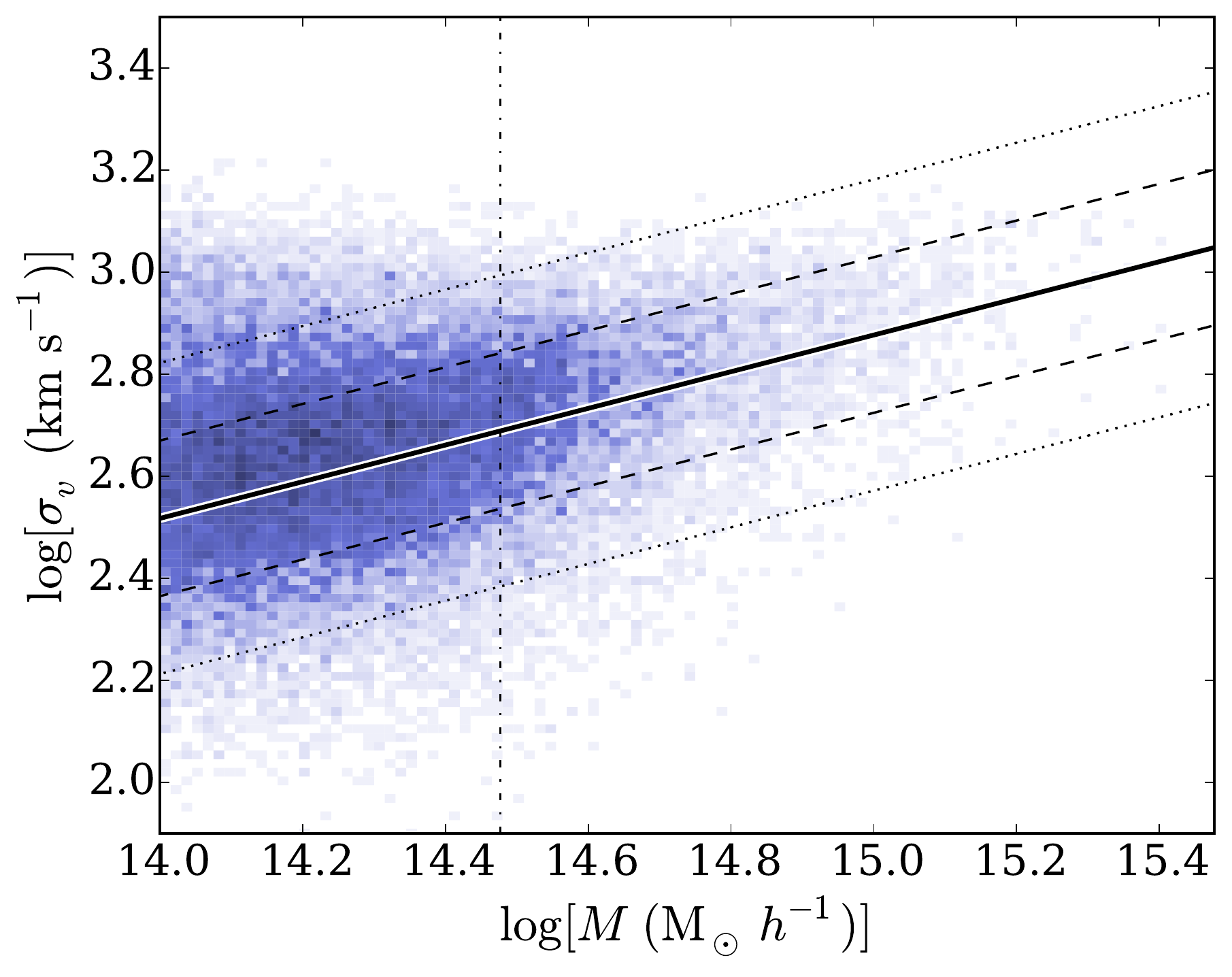}
  \caption{Velocity dispersion, $\sigma_v$, vs.~cluster mass, $M$, for a simple cylindrical cut with iterative 2-$\sigma$ paring.  Clusters above $3\times10^{14}\,\Msolarh$ (vertical black dash dotted) inform the fit (black solid) and determine the lognormal scatter (68\% and 95\%, dashed and dotted, respectively).  The presence of interlopers introduces significant scatter, particularly at low masses, where the effect of interlopers is more pronounced.} 

  \label{fig:msig}
\end{figure}
 
 In a typical power-law scaling relation, one starts with the virial theorem to find a relationship between the velocity dispersion, $\sigma_v$, and halo mass, $M$.  This power law is given as $\sigma_v \propto M^{1/3}$, but can be rewritten more generally as
\begin{equation}
\sigma_{v}(M) = \sigma_{15}\left( \frac{ M}{10^{15} \, \Msolarh} \right)^\alpha.
\label{eq:powerlaw}
\end{equation}
where $\sigma_{15}$ is the typical velocity dispersion of galaxies residing within a $10^{15}\, \Msolarh$ halo and the parameter $\alpha$ is allowed to vary from the theoretically-predicted $\alpha=1/3$ and is instead fit to data.  The best fit is then be used to predict cluster mass from a velocity dispersion of galaxies.  When applied to the \memb{} Catalog, this method will be denoted \PLM{}, and when applied to the \interloper{} Catalog, it will be denoted \PLI{}.

To account for a potentially-changing slope caused by the cylindrical cut used for the \interloper{} Catalog, a lower mass cut of $3\times10^{14}\Msolarh$ will be applied to the data used to fit the power law.  We find a least-squares fit to $\log(\sigma_v)=\alpha\log(M)+\beta$ for the PL Train Catalog.  

While \PLM{} is well-described by $\alpha=0.382$, $\sigma_{15}=1244\,\kms$, \PLI{} has a shallower slope and smaller velocity dispersion expected for a $10^{15}\,\Msolarh$ halo, $\alpha=0.359$ and $\sigma_{15}=753\,\kms$, respectively.  These best fit parameters to the $M(\sigma_v)$ power law (Equation \ref{eq:powerlaw}) for each catalog are noted in Table \ref{table:catalog}.  The scaling relation best fit for the \interloper{} Catalog is shallower and has a smaller $\sigma_{15}$ compared to that of the \memb{} Catalog, therefore, applying the \PLM{} fit to observed clusters with interlopers can introduce additional error.  We additionally caution that these parameters are a fit for a particular simulation and cylindrical cut and should be applied to observational data with care.  

The introduction of interlopers is a large source of scatter in $M(\sigma_v)$.  Figure \ref{fig:msig} shows a two-dimensional histogram of $\sigma_v$ vs.~$M$ for the \interloper{} Catalog.  Overlaid is a best fit with 1- and 2-$\sigma$ lognormal errors calculated for clusters with mass above $3\times10^{14}\, \Msolarh$ and extrapolated down to lower masses.  This lognormal scatter, $\sigma_\mathrm{gauss}$, is determined by the standard deviation of the residual, $\delta$, defined as
\begin{equation}
	\delta = \log(\sigma_\mathrm{measured})-\log(\sigma_\mathrm{expected}).
	\label{eq:residual}
\end{equation}
Here, $\sigma_\mathrm{measured}$ is the velocity dispersion of the galaxies within the pared cylinder and $\sigma_\mathrm{expected}$ is the typical velocity dispersion expected for a cluster of a given mass{, found by applying Equation \ref{eq:powerlaw} with true cluster mass $M$ and best fit parameters $\sigma_{15}$ and $\alpha$}.  Of halos with $M\geq3\times10^{14}\,\Msolarh$, 1\% reside above the $+2\sigma$ dotted line and 4\% reside below the $-2\sigma$ dotted line.  However, of halos with $1\times10^{14}\, \Msolarh \leq M < 3\times10^{14}\,\Msolarh$, 8\% reside above $+2\sigma$ and 4\% below $-2\sigma$.  The scatter found for the higher-mass clusters is clearly not descriptive of the lower-mass clusters; this is explored further in the Appendix (Sec.~\ref{sec:appendix}).
 
 The \PLM{} and \PLI{} approaches rely on a single summary statistic, $\sigma_v$, to describe the dynamics of the cluster members.  However, mergers and infalling matter, for example, can distort the shape of the velocity PDF and cause the cluster's mass to be overpredicted by a traditional power-law approach.  Next, we will explore a machine learning approach for predicting cluster masses that learns from a distribution, rather than from a single summary statistic.

\begin{deluxetable*}{l l l r r l}
\tabletypesize{\scriptsize}
\tablecaption{Feature Summary \label{table:featsummary}}
\tablewidth{0pt}
\tablehead{
\colhead{Case} & \colhead{Approach} & \colhead{Train \& Test Catalogs} & \colhead{Summary Stats} & \colhead{Distribution Features} &
\colhead{Color} 
}
\startdata
\PLM{} 	& Power Law 				&\memb{} 					& $\sigma_v$ 			&---							& Red\\ 
\PLI{}  	& Power Law  				&\interloper{} 				& $\sigma_v$ 			&---							& Blue \\ 
\MLv{}			& Machine Learning: SDM  	&\interloper{} 			 	&---					& $|\vlos|$  					& Green \\ 
\MLR{} 			&  Machine Learning: SDM 	&\interloper{} 			 	&---					& $R$ 						& Orange \\ 
\MLvR{}			&  Machine Learning: SDM 	&\interloper{}				&---					& $|\vlos|$ \& $R$				& Brown \\ 
\MLvsR{} 			&  Machine Learning: SDM  	&\interloper{}			 	&---					& $|\vlos|$, $\vsig$, \& $R$ 		& Purple \\ 
\enddata

\end{deluxetable*}

\subsection{{Support Distribution Machines}}
\label{sec:SDM}

Support distribution machines \citep[SDMs;][]{2012arXiv1202.0302S} are a class of machine learning algorithms built upon Support Vector Machines \citep[SVMs;][]{Drucker97supportvector,scholkopf2002learning}.  Given a training set of (distribution, scalar) pairs, the goal of SDM is to learn a function that predicts a scalar from a distribution.  They will be applied here to learn from distributions of galaxy observables such as galaxy LOS velocity and projected distance from cluster center.  These distributions of galaxy observables will then be implemented to predict the log of the cluster mass, $\log(M)$.  

The SDM method applied requires the divergence between pairs of distributions in the training and test sets. For this purpose, we employ the Kullback-Leibler (KL) divergence, and estimate the divergence via the estimator from  \cite{4839047}. This is a k-nearest-neighbor-based estimator. In practice, we use k=3. The relative divergences from training data are used to select SDM best fit kernel parameters $C$ and $\sigma$, the loss function parameter and Gaussian kernel parameter, respectively, via 3-fold cross-validation.  These are used to train the regression model with the selected best-fit kernel, which in turn is used to predict masses for the test data.  For a full discussion of SVM formalism as well as a discussion of how SDM deviates from the SVM base case, see \cite{2012arXiv1202.0302S} and \cite{Ntampaka2015}. 

In order to take full advantage of the available data, we cyclically learn from 90\% of the clusters and predict masses from the remaining, independent 10\%; this is repeated ten times until the masses of all clusters in the \interloper{} Catalog have been predicted.  To prepare the mock cluster catalog for SDM implementation, clusters are rank-ordered by mass and sequentially assigned to one of ten folds.  Multiple LOS views of a unique cluster are all assigned to the same fold, ensuring that each time SDM is implemented, a unique cluster is used either for training or for predicting, but never both.

Of the ten folds, nine from the \interloper{} Train Catalog are used to select SDM best fit kernel parameters $C$ and $\sigma$ and subsequently train the regression model with the selected kernel.  This regression model is then used to predict the masses of the clusters in the tenth fold of the \interloper{} Test Catalog.  The process is repeated ten times, training on nine Train Catalog folds and predicting the tenth Test Catalog fold, until masses for the entire \interloper{} Test Catalog have been predicted.

 We implement SDM with four sets of {galaxy features: the PDF of galaxy LOS absolute velocity ($|\vlos|$), the PDF of normalized velocity ($|\vlos|/\sigma_v$), the PDF of projected distance from the cluster center ($R$), and combinations thereof}.  As discussed in \cite{Ntampaka2015}, features must be chosen with care because features uncorrelated with mass tend to wash out the effects of the more important features.  The motivation for features implemented here is as follows:

 \begin{figure}[]
  \centering
  \includegraphics[width=0.5\textwidth]{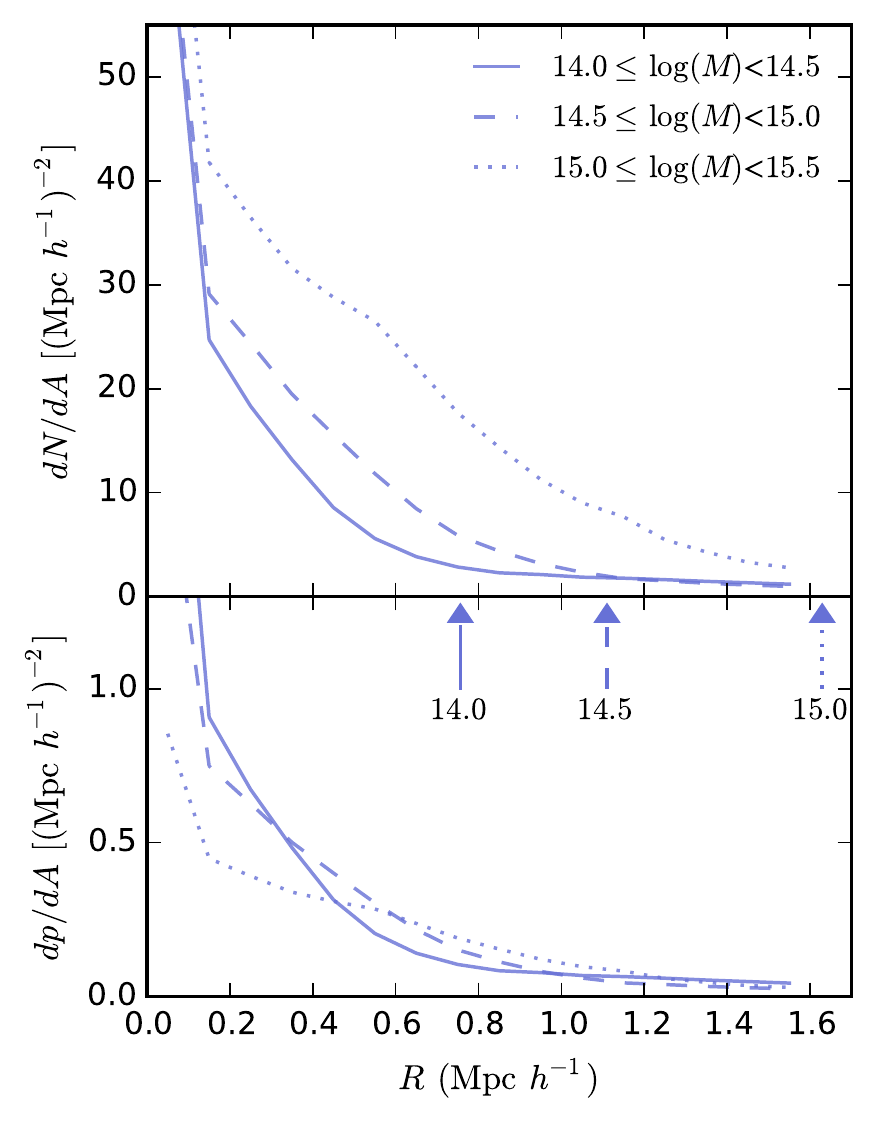}
  \caption{Top:  Average number of galaxies per unit plane-of-sky area, $dN/dA$, vs.~projected distance from the center of the cluster, $R$, for three $\log[M(\Msolarh)]$ ranges in the \interloper{} Test Catalog, in 0.1 $\Mpch$ bins.  The shape and amplitude of this effective column density vary with the mass of the primary halo.  Bottom:  Probability of finding a galaxy per unit area, $dp/dA$, vs.~$R$.   The shape and amplitude of this measure also varies with primary halo mass.  Arrows denote the characteristic radius of a halo with $\log[M(\Msolarh)]$ indicated.  SDM trained on the feature $R$ takes advantage of how the {distribution of subhalo radius} changes with mass to predict a halo mass based on the distribution of $R$.} 

  \label{fig:Rstack}
\end{figure}

 \begin{enumerate}

\item \MLv{}: The use of velocities is motivated by the virial theorem, as we have seen in Figure \ref{fig:msig} that velocity dispersion of galaxies, $\sigma_v$, relates to mass as a power law, albeit with significant scatter.  The \MLv{} catalog uses absolute value of galaxy LOS velocities, $|\vlos|$, as a single feature for training and testing by means of SDM.  

\item \MLR{}:  Even in the presence of interlopers, galaxy density profiles can be used to determine cluster mass  \citep[e.g.][]{2005ApJ...633..122H, 2015MNRAS.449.3082P}.    This is motivated by Figure \ref{fig:Rstack}, which shows stacked halos from the \interloper{} Test Catalog divided into three $\log[M(\Msolarh)]$ bins.  Despite the fixed aperture, the number of galaxies per unit plane-of-sky area ($dN/dA$) in concentric rings has a markedly different distribution for the low, middle, and high-mass halos.  The probability of finding a galaxy per unit plane-of-sky area ($dp/dA$) also exhibits a unique shape for each mass bin.  For this reason, we will consider an \MLR{}  catalog, with the galaxy radii from the halo center, $R$, as the sole feature.
 
\item \MLvR{}: Decreasing velocity dispersion profiles have been noted in clusters \citep[e.g.][]{2003AJ....126.2152R}.  Because $\vlos$ and $R$ individually can provide information about cluster mass, it seems reasonable that the joint probability distribution of $|\vlos|$ and $R$ may be informative as well.    \MLvR{} will learn from the joint distribution of the LOS velocity feature, $|\vlos|$, and the galaxy radius feature, $R$, in a two-dimensional feature space.  

\item \MLvsR{}: The shape of the velocity PDF can be indicative of mass accretion and mergers \citep{Evrard:2008aa, 2011MNRAS.413L..81R}.  As found in \cite{Ntampaka2015}, explicitly normalizing $\vlos$ by its width, $\sigma_v$, can emphasize these shape differences and improve mass predictions, particularly at the high-mass end.  We will consider a training set, \MLvsR{}, that employs $|\vlos|$, $|\vlos|/\sigma_v$, and $R$ in a three-dimensional features space.
 
\end{enumerate}
These ML method names and corresponding distribution features are summarized in Table \ref{table:featsummary} for reference and will be used by SDM to predict cluster masses.  Next, we will explore how the PL's scaling relation and ML's distribution-to-scalar approach predicted masses of clusters from the mock cluster catalog.

\section{Results}
\label{sec:results}

\subsection{Power Law }
\label{sec:PLresults}

 \begin{figure*}[!t]
\begin{center}
\begin{tabular}{c c}
        	\includegraphics[width=0.45\textwidth]{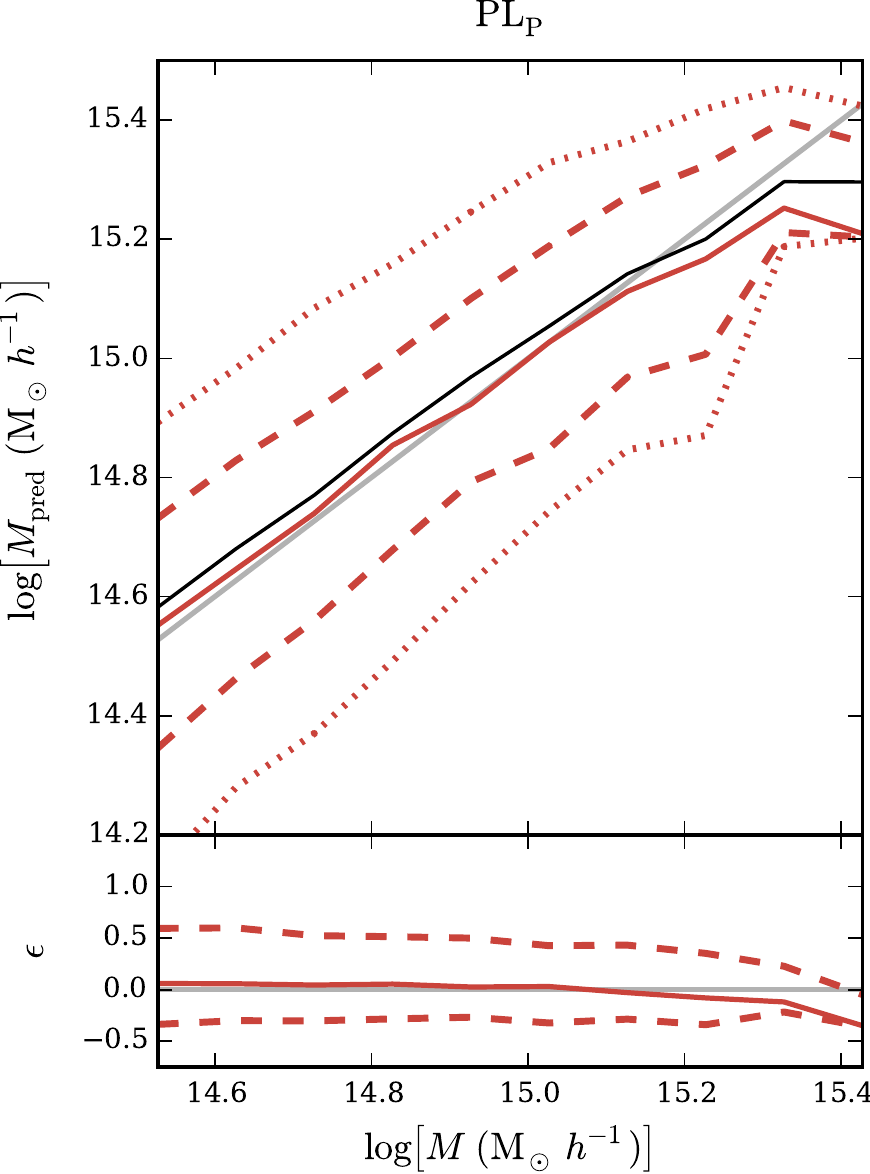} &\includegraphics[width=0.45\textwidth]{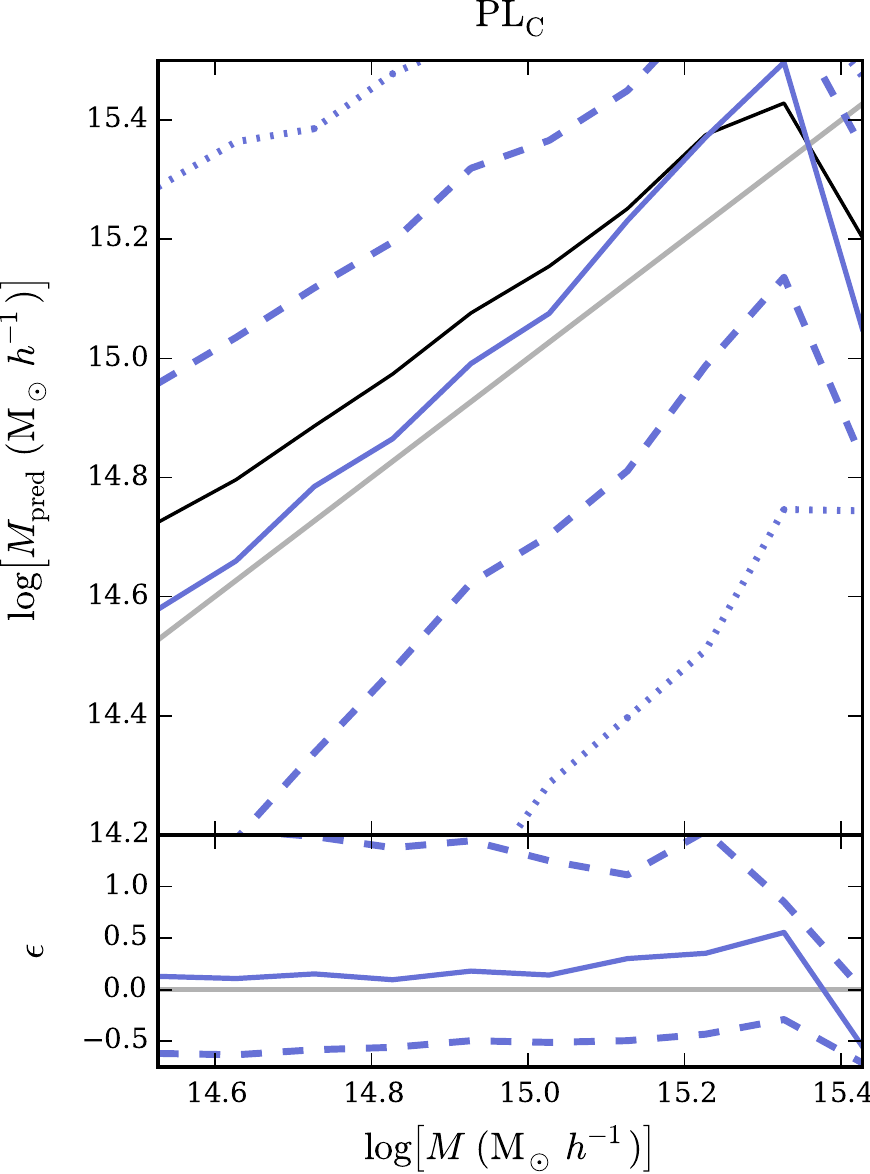} \\
\end{tabular}
      	 \caption{Left:   Power-law scaling relation applied to the \memb{} Catalog (method \PLM{}).  Predicted vs.~true mass, binned in 0.1 $dex$ $\log[M \, (\Msolarh)]$ bins, with mean (black solid), median (red solid), 68\% (dashed), and 95\% (dotted) scatter shows that significant scatter exists even when applying a scaling relation to a catalog of pure and complete clusters (top).  Though the mass error median (red solid)  is nearly zero (gray solid), it has significant 68\% scatter (red dashed) (bottom).  	 Right:  Power law scaling relation applied to the \interloper{} Catalog, which contains impure and incomplete clusters (method \PLI{}).   The imperfect catalog introduces additional scatter in $\epsilon$ compared to the \PLM{} case, most notably at low masses where the sample impurity is particularly pronounced.  These two plots provide best (left) and worst (right) case scenario benchmarks for applying an $M(\sigma_v)$ power law scaling relation to cluster observation.}
       	\label{fig:PL}
      	\end{center}
\end{figure*}

Figure \ref{fig:PL} shows the predicted vs.~true cluster masses for the \memb{} and \interloper{} Catalogs.   When a power law is applied to the \memb{} Catalog, there is significant scatter in mass predictions.  The bottom panel of Figure \ref{fig:PL} shows the median and 68\% scatter in the fractional mass error, $\epsilon$, given by
\begin{equation}
	\epsilon=(M_{\textrm{pred}}-M)/M,
	\label{eq:fracerror}
\end{equation}
where $M$ is the true cluster mass and $M_\textrm{pred}$ is the predicted cluster mass.  The scatter in \PLM{} errors can be attributed to both {physical and selection} effects.  For example, infalling matter tends to create a velocity PDF with negative kurtosis, tending to overpredict the mass.  Cluster mergers \citep{Evrard:2008aa}, galaxy selection effects \citep{2013ApJ...772...47S}, and dynamical friction and tidal disruption \citep{2013MNRAS.430.2638M} can each play a role in contributing to this scatter.

Figure \ref{fig:PL} also shows results for the power-law scaling relation applied to the \interloper{} Catalog.  Impure and incomplete clusters introduce further scatter and errors increase significantly.  This scatter is most notable at the low-mass end, where the inclusion of interlopers is most prominent.

\PLM{} and \PLI{} serve as upper and lower bounds for errors for a power-law scaling relation: \PLM{}'s pure and complete clusters show the level of scatter that remains when interlopers are completely eliminated, while \PLI{}'s simplistic interloper removal technique highlights how interlopers can affect scatter in an extreme case.  More effective interloper removal methods are available, applying more discriminating statistical techniques \citep[e.g.][]{1996ApJ...473..670F, 2007MNRAS.379..867V,2013MNRAS.429.3079M}, with some considering only red elliptical galaxies which preferentially reside in clusters \citep[e.g.][]{2013ApJ...772...47S}.  We expect a more refined interloper-removal scheme to reside between the two benchmark cases shown in Figure \ref{fig:PL}.  

One may consider the possibility of improving mass predictions by extending mass range for training.  However, due to the existence of many high-error, high-$\sigma_v$ clusters shown in Figure \ref{fig:msig}, decreasing the lower mass limit may not improve mass predictions.  Even without this high-error population, the power law dynamical mass approach has significant scatter exacerbated by the presence of interlopers.  Further, the potentially informative infalling galaxy observations have not been considered, nor have the baseline LOS velocity PDF shapes indicative of a nonvirialized or merging system.  Next, we will explore the results of learning on full distributions with a machine learning approach.

\begin{figure*}[!h]
\begin{center}
\begin{tabular}{c c}
     \includegraphics[width=0.4\textwidth]{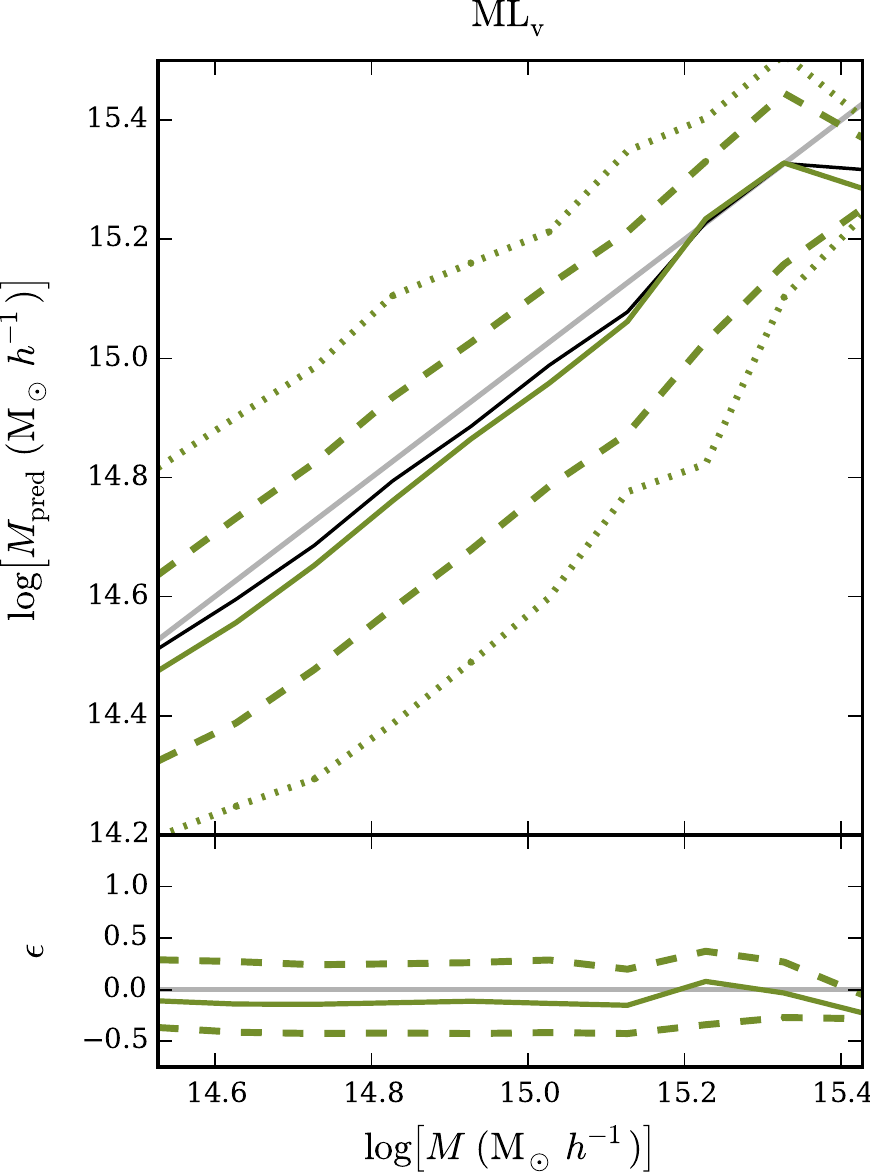} & \includegraphics[width=0.4\textwidth]{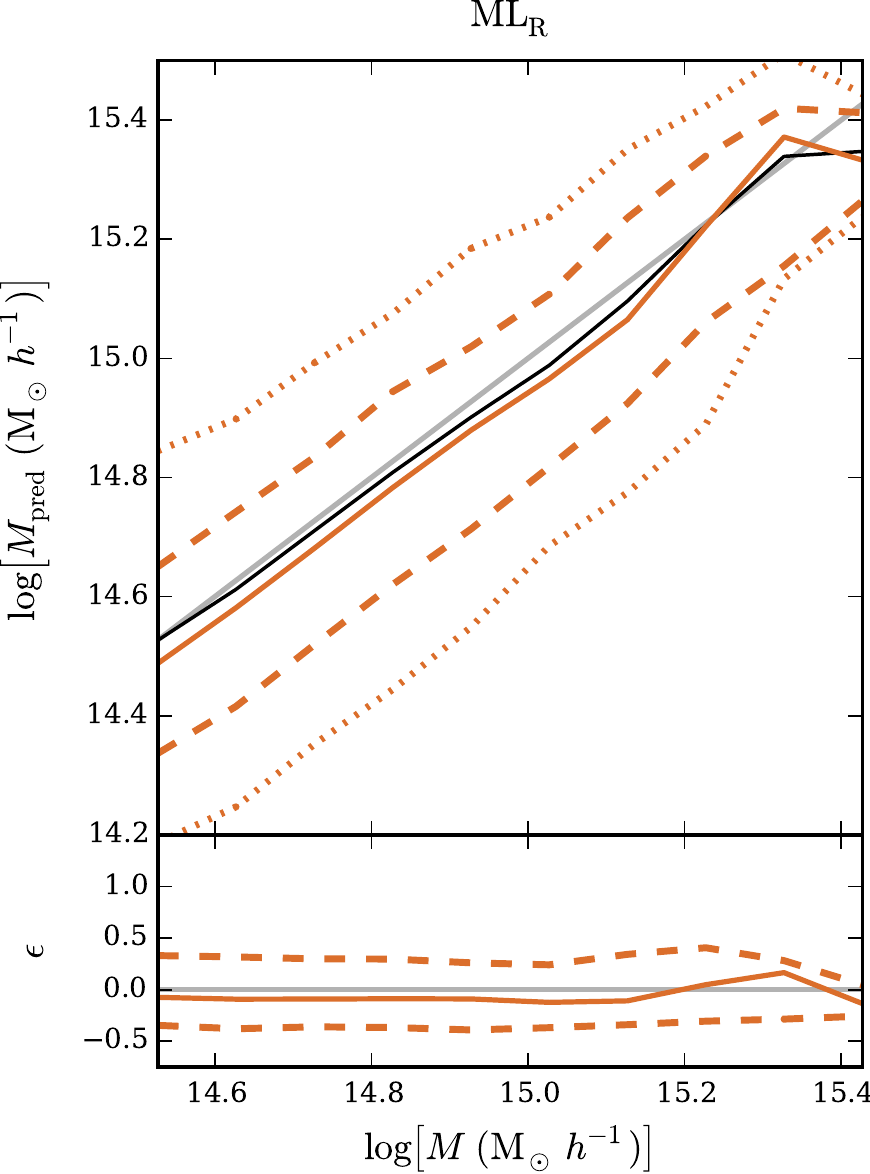} \\ [1.5mm]
     \\
     \includegraphics[width=0.4\textwidth]{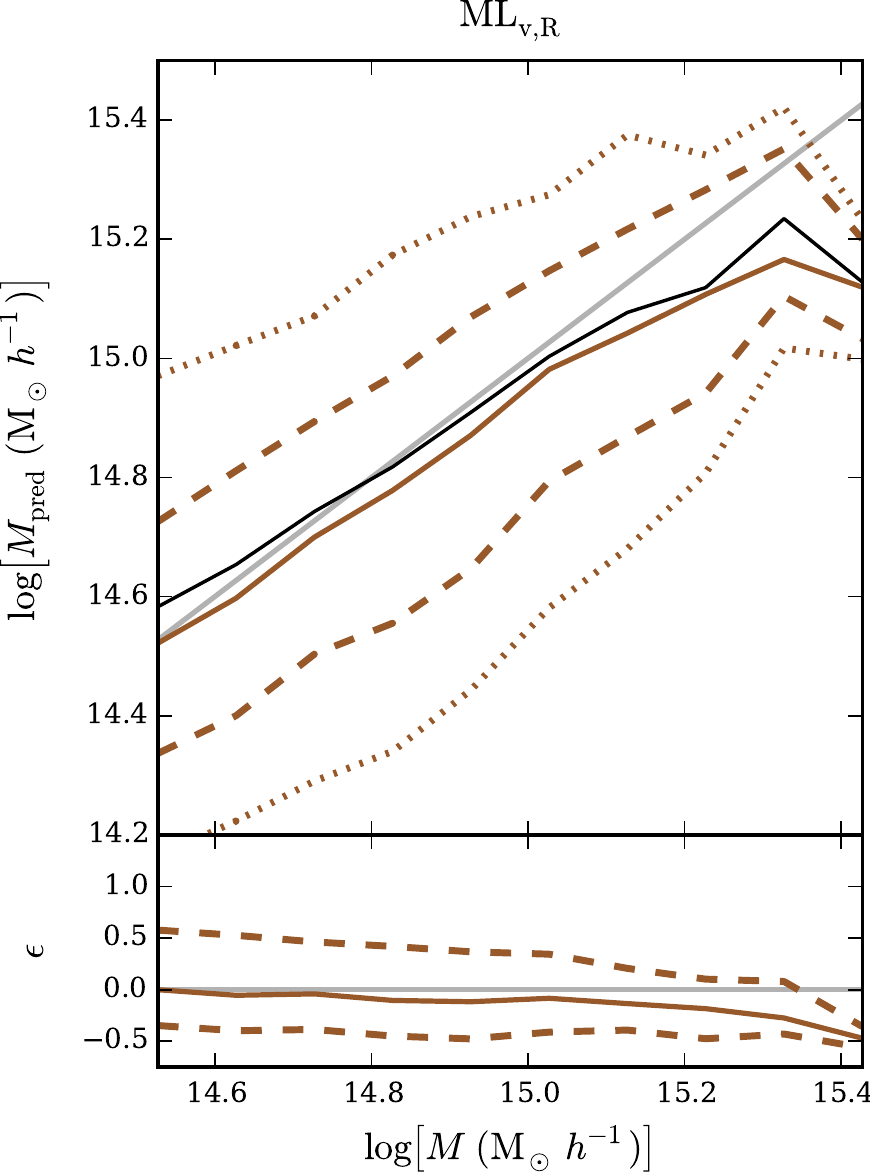} & \includegraphics[width=0.4\textwidth]{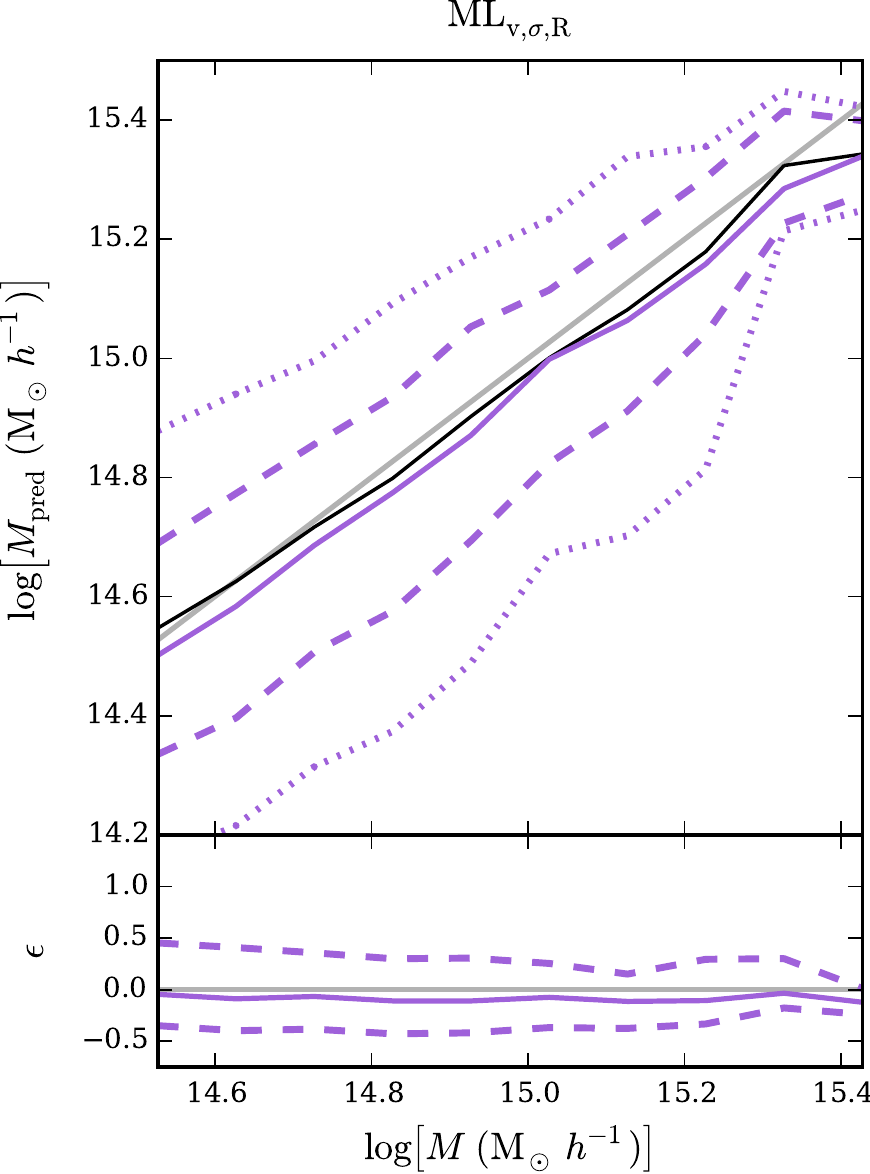} \\ [1.5mm]

  \end{tabular}
      \caption{\small{Top Left:  SDM results for \MLv{} (green).  The predicted vs.~true mass is binned in 0.1 $dex$ $\log[M(\Msolarh)]$ bins.  Mean (black solid), median (colored solid), 68\% (dashed), and 95\% (dotted) scatter are shown (top).  The median error (solid) and error 68\% scatter (dashed) are also shown (bottom).  \MLv{} gives better than a factor-of-two reduction in the width of error compared to a standard scaling relation applied to the same catalog.
      	Top Right: SDM results for \MLR{} (orange).  \MLR{} and \MLv{} minimize the width of the error distribution.
	Bottom Left:  SDM results for \MLvR{} (brown).  \MLvR{} underpredicts at high masses and is therefore identified as a disfavored method.
	Bottom Right:  SDM results for \MLvsR{} (purple).  \MLvsR{} minimizes the tendency to underpredict across mass range. }
	}
      \label{fig:MLsummary}
      \end{center}
\end{figure*}

\begin{figure*}[t]
  \centering
  \includegraphics[width=\textwidth]{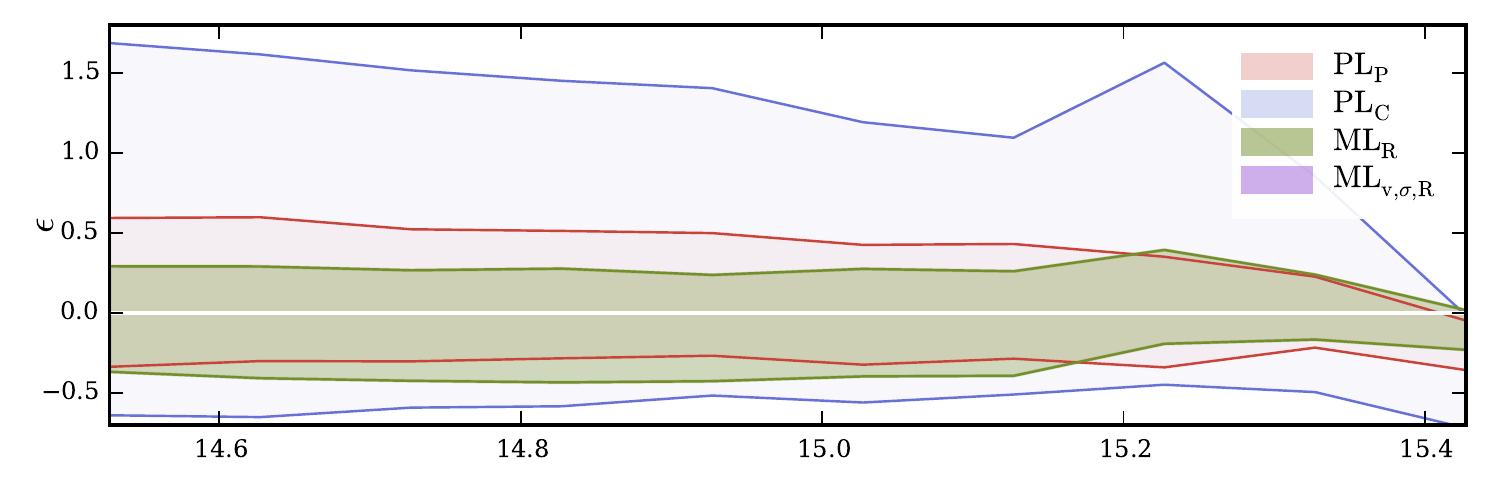}
  \includegraphics[width=\textwidth]{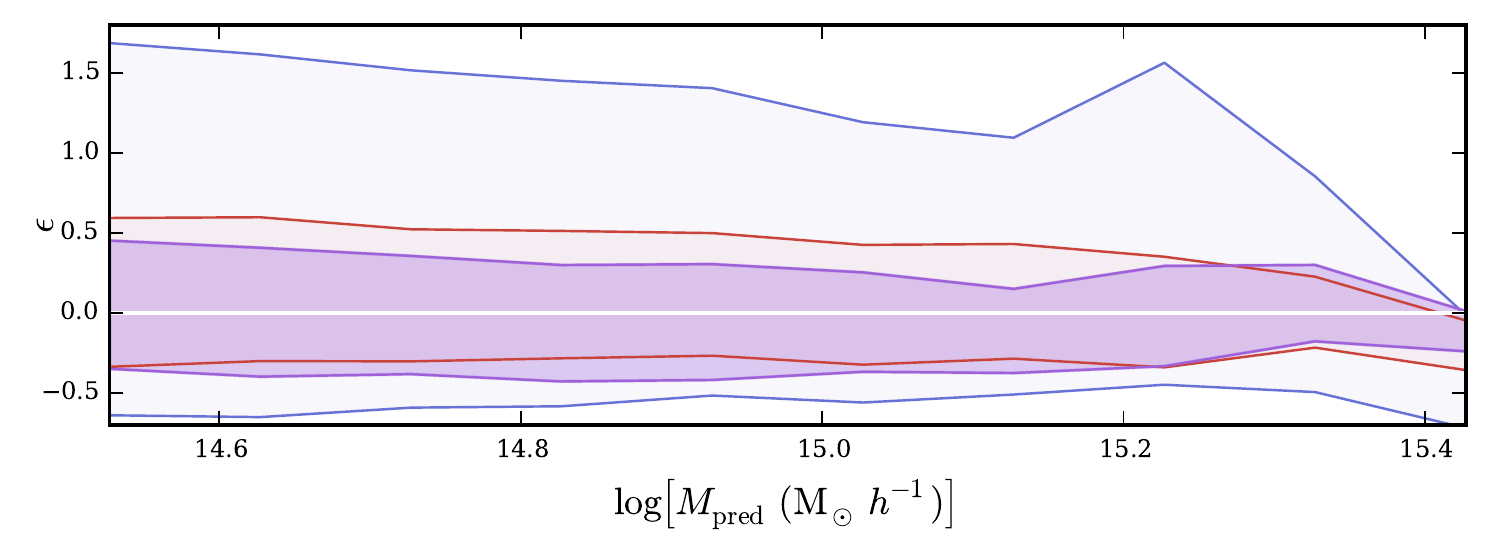}
  \caption{Top:  Error 16th and 84th percentiles (i.e.~68\% scatter) as a function of mass for \MLv{} (green) as compared to a power-law approach applied to the \memb{} Catalog (\PLM{}, red) and to the \interloper{} Catalog (\PLI{}, blue).  Bottom:  Error scatter as a function of mass for \MLvsR{} (purple) compared to \PLM{} and \PLI{}.  The errors of a dynamical mass power-law approach with a more refined interloper removal scheme should be bounded by \PLI{} and \PLM{}.  However, even when trained on the impure and incomplete catalog that produced the blue \PLI{} results, \MLv{} and \MLvsR{} have $\epsilon$ width comparable to or smaller than the best case \PLM{} power law.\\ \\ \\}
  \label{fig:err}
 \end{figure*}

\begin{figure*}[!tbh]
\begin{center}
\begin{tabular}{c c c}

     \includegraphics[width=0.3\textwidth]{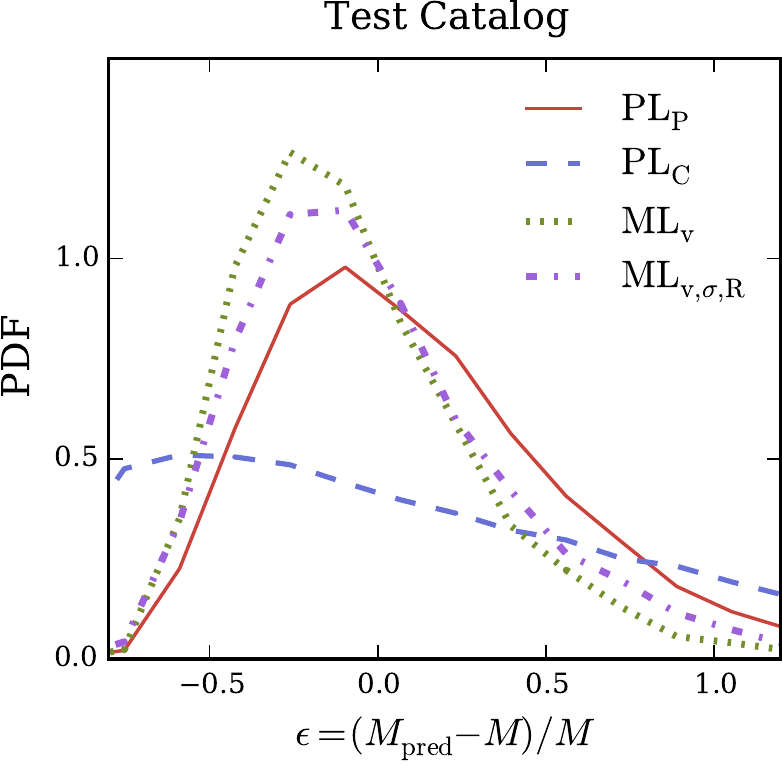} & \includegraphics[width=.3\textwidth]{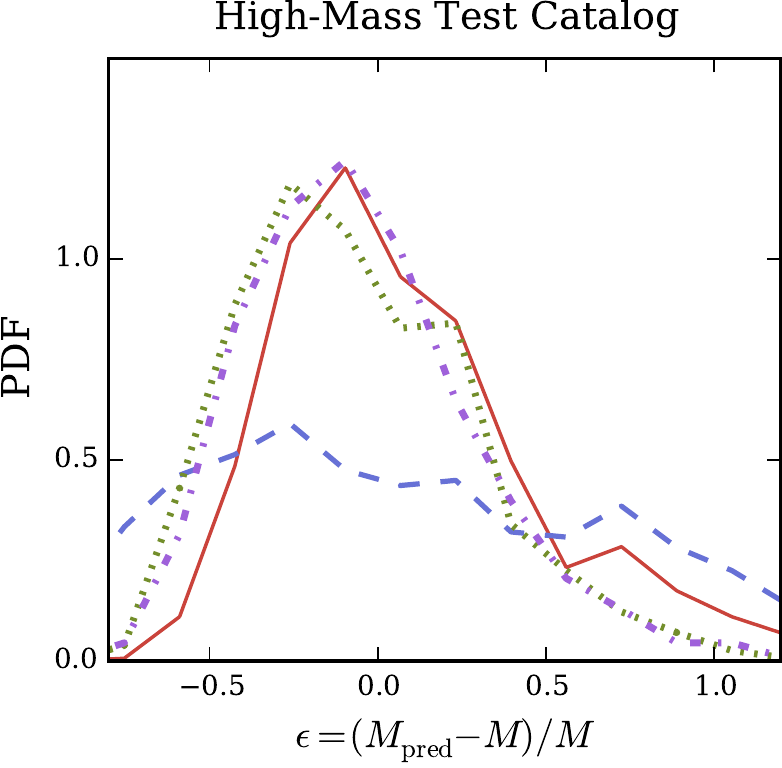}& \includegraphics[width=.3\textwidth]{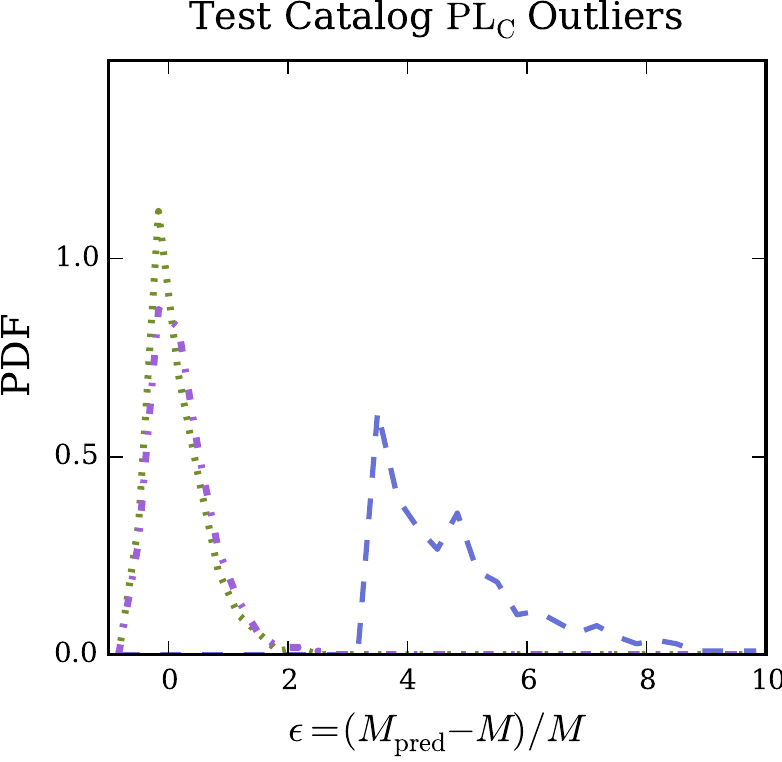} \\
  \end{tabular}
       \caption{Left:  PDF of fractional mass errors for the Full Test Catalogs.  A power-law $M(\sigma_v)$ scaling relation for a catalog of pure and complete clusters shows significant errors (\PLM{}, red solid).  The error distribution widens further when interlopers contaminate the clusters (\PLI{}, blue dashed).  Remarkably, SDM (\MLv{}, green dotted, and \MLvsR{}, purple dash dotted) applied to the \interloper{} Catalog outperform the $M(\sigma_v)$ scaling relation applied to the \memb{} Catalog.  Center:  PDF of errors for the High-Mass Test Catalogs ($M\geq7\times10^{14}\, \Msolarh$) shows a similar trend for rare, high-mass halos; the ML approaches minimize error significantly over a power-law scaling relation applied to the same catalog.  Right:  PDF of the high-$\delta$, high-\PLI-error population of clusters.  While the power law catastrophically overestimates the masses of these outlying objects, ML approaches perform well, with a PDF of fractional mass errors for these outliers that is only slightly wider than is found for the full catalog. \\ \\}
      \label{fig:PDF}
      \end{center}
\end{figure*}

\begin{figure*}[!t]
  \centering
  \includegraphics[width=\textwidth]{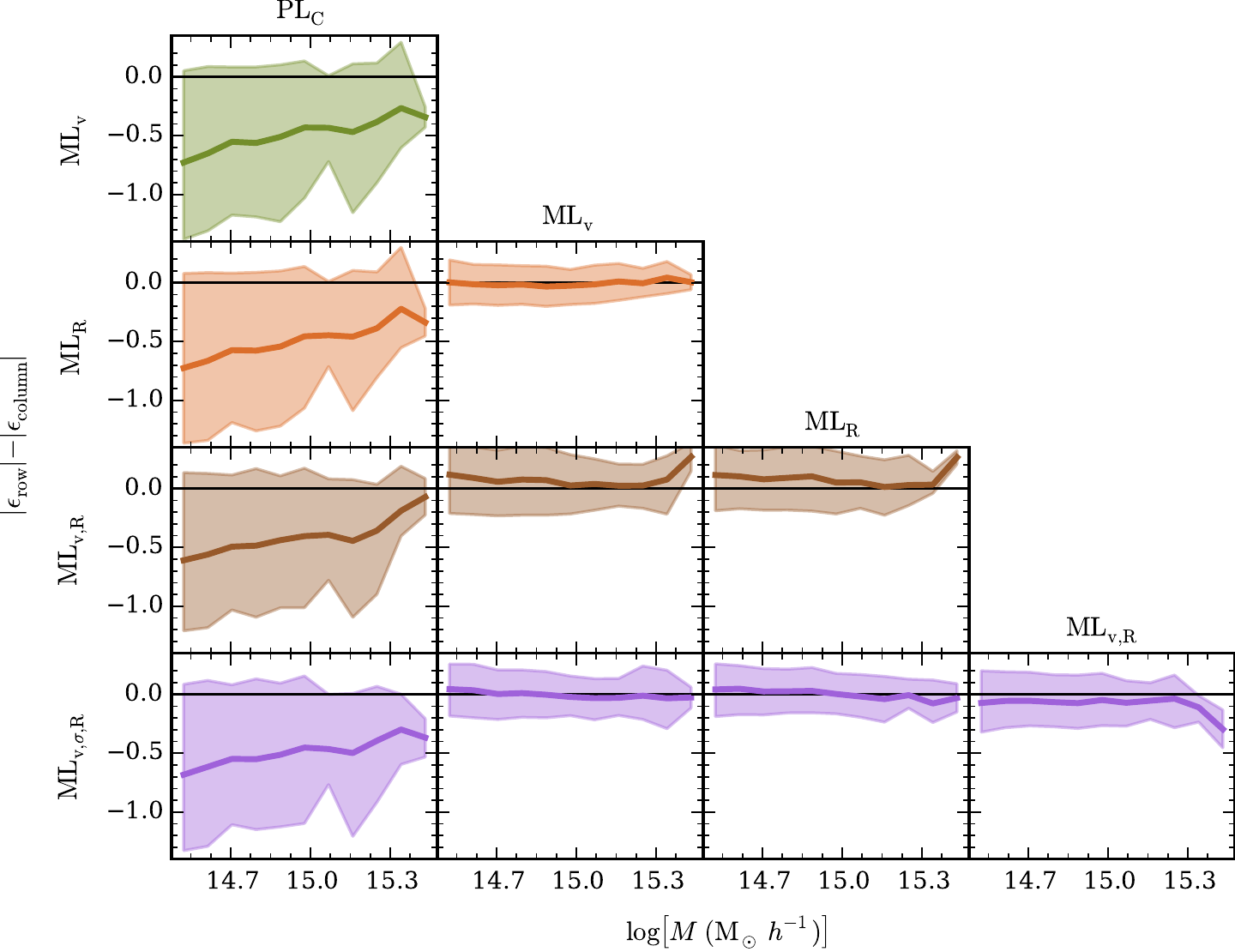}
  \caption{Summary comparison of the five methods trained and tested on the \interloper{} Catalog, with difference in absolute error, $|\epsilon_{\mathrm{row}}|-|\epsilon_{\mathrm{column}}|$, as a function of mass (see Equation \ref{eq:fracerror}).  Values below the solid black $0$ line indicate that the row method is performing better than the column method for a given mass bin.  The left column summarizes a comparison of the four new SDM methods to the \PLI{} power law; SDM with any of the four feature combinations improves mass predictions in all mass bins.  While \MLvR{} outperforms \PLI{}, it performs poorly at high masses compared to the other ML methods.}
  \label{fig:waterfall}
\end{figure*}

\subsection{Machine Learning }
\label{sec:MLresults}

Figure \ref{fig:MLsummary} shows the SDM predictions for each of the four feature sets:  \MLv{}, \MLR{}, \MLvR{}, and \MLvsR{}.  As in Figure \ref{fig:PL}, the top panel shows predicted vs.~true mass median with 68\% and 95\% scatter.  Each of the ML methods reduces scatter significantly compared to \PLI{}, the power law that is applied to the same catalog as these ML methods.  One should not overly interpret the fluctuations in the two largest mass bins, as they contain only six unique clusters, a small fraction of the total clusters in the sample.  The bottom panel shows median error $\epsilon$ (see Equation \ref{eq:fracerror}) with 68\% scatter.  The 68\% scatter is dramatically reduced compared to the power law relation with the same catalog, \PLI{}, and is comparable to the power law relation with a catalog of pure and complete clusters, \PLM{}.  \MLvsR{} has median binned mass predictions that are closest to the true mass, while \MLR{} has the smallest error width, but all four ML methods outperform \PLI{} by a large margin.

A comparison of mass predictions is presented in Figure \ref{fig:err}.  PL provides two benchmarks: while the \PLI{} error shows what we might expect from a impure and incomplete interloper catalog, \PLM{} gives a best-case scenario where cluster members are perfectly known and interlopers are entirely excluded.  Across the entire mass range considered, \MLv{} and \MLvsR{} exhibit a dramatically tighter error distribution than a power law applied to the \interloper{} Catalog.  Even in comparison to the \memb{} Catalog, SDM produces a tighter error distribution.

Figure \ref{fig:PDF} shows a PDF of errors for all clusters above $3\times10^{14}\,\Msolarh$ and for those above $7\times10^{14}\,\Msolarh$.  The  \PLI{} curve shows the PDF of errors associated with $M(\sigma_v)$ power law with the \interloper{} Catalog's simple cylindrical cut about cluster centers.  In contrast, the \PLM{} curve shows the PDF of erros associated with the $M(\sigma_v)$ power law of the \memb{} Catalog,  built from perfect knowledge of cluster members.  For both \MLv{} and \MLvsR{}, the number of extreme overpredicted masses with $\epsilon \gtrsim 0.6$ is dramatically reduced over even the \PLM{} power law.  The extreme underpredicted masses with $\epsilon \lesssim -0.6$ are reduced compared to \PLI{}. 

The mean error ($\mean{\epsilon}$) and median with central 68\% width ($\epsilon \pm \Delta \epsilon$) of these PDFs are summarized in Table \ref{table:methodcomp}.  Here we see PL's tendency to overpredict (positive $\epsilon$ and $\mean{\epsilon}$) in contrast with ML's tendency to underpredict (negative $\epsilon$ and $\mean{\epsilon}$).  ML's underpredictions are caused by the hard upper mass limit and dearth of unique training halos at the high-mass end.  The resulting underprediction is most conspicuous in \MLv{} (both the \interloper{} Test and \interloper{} High-Mass Test) and in \MLvR{} (\interloper{} High-Mass Test only).  \MLvR{} has the smallest error offset (-0.04), but does so at the cost of underpredicting the highest-mass clusters. This bias is most evident at the higher mass end, where halos' masses are systematically underpredicted.  Because of this pronounced bias, \MLvR{} is therefore identified as a disfavored method.

The relative error widths ($\Delta \epsilon$) for all ML methods for all methods are more than a factor-of-two smaller than \PLI{} (69\%, 69\%, 58\%, and 64\% for \MLv{}, \MLR{}, \MLvR{} and \MLvsR{}, respectively).  Even compared to \PLM{} which is applied to the \memb{} Catalog, SDM produces a smaller relative error width (23\%, 23\%, 3\%, and 12\% for \MLv{}, \MLR{}, \MLvR{} and \MLvsR{}, respectively).

As we saw in Figures \ref{fig:msig} and \ref{fig:PL}, there is a wide scatter in $\sigma_v$ associated with the \interloper{} Test Catalog.  Shown in the right panel of Figure \ref{fig:PDF} are the clusters for which \PLI{} severely overestimated cluster mass.  These objects are particularly worrisome, as are predicted by \PLI{} as being much more massive than they truly are, appearing to be rare, high-mass clusters.  These outliers are isolated by their residual, $\delta$ (Equation \ref{eq:residual}); each has $\delta \geq 1.5\times\sigma_\mathrm{gauss}$.  We find the ML error PDF for these objects is centered on zero, with a PDF width only slightly wider than the one shown in the left panel of Figure \ref{fig:PDF} for the full catalog.  Further, while the \PLI{} method overpredicts catastrophically, the ML methods predict much more reasonable masses.

Figure \ref{fig:waterfall} shows a comparison of the five methods applied to the \interloper{} Catalog: \PLI{}, \MLv{}, \MLR{}, \MLvR{}, and \MLvsR{}.  The difference in absolute errors, denoted $|\epsilon_\mathrm{row}|-|\epsilon_\mathrm{column}|$, gives a measure of how well the row method predicts compared to the column method; values below $0$ are indicative of the row method predicting more accurately.  The left column of this plot shows a comparison of ML to PL; all four ML methods consistently predict masses with a much smaller error than \PLI{}.  The mean difference in absolute value of errors, denoted $|\epsilon|-|\epsilon_{\mathrm{PL_I}}|$, is summarized in Table \ref{table:methodcomp}.  This summary statistic quantifies the mean value shown in the left column of Figure \ref{fig:waterfall}.  The more negative this value, the more reduced a model's errors compared to \PLI{}.  Model \MLR{} decreases error $\epsilon$ by an average of 0.61 compared to \PLI{}; \MLR{} is the best ML method by this measure.  The right three columns of Figure \ref{fig:waterfall} compare the ML techniques to one another. \MLvR{} is shown here to be the weakest of the ML methods; though it outperforms \PLI{} by a large margin, SDM produces more accurate mass predictions when applied with other feature sets.

As in \cite{Ntampaka2015}, pairing $|\vlos|$ with the feature $|\vlos|/\sigma_v$ accentuates differences in velocity PDF shape and highlights, for example, the wide, flat hallmark PDF of a halo experiencing infalling matter.  As a result of this additional feature, the mean and median errors edge closer to the desired values of zero.  This offers an explanation as to why the three-feature set of \MLvsR{}  shows a mean error closer to zero (0.01) compared to \MLv{} and \MLR{}.  \MLvsR{} is identified as the preferred feature set for minimizing error bias.

Though \MLvR{} employs two features that are highly-correlated with mass, these features reside in a two-dimensional feature space.  The joint distribution of $|\vlos|$ and $R$ is likely too sparsely sampled by the galaxies in an individual cluster to make a strong correlation between this joint distribution and cluster mass.  This effect becomes particularly pronounced for rare, massive clusters, which are underpredicted by \MLvR{}.  

\MLvsR{}, however, predicts the masses of these clusters well.  This may be explained by the nature of the third feature, $|\vlos|/\sigma_v$.  Though the probability distribution employed by \MLvsR{} resides in a three-dimensional feature space, the combination of $|\vlos|$ with $|\vlos|/\sigma_v$ constrains individual clusters' distributions to lie on a plane.  These planes are sorted in the three-dimensional space by their slope, $\sigma_v$.  This sorting effectively isolates high-$\sigma_v$ clusters from low-$\sigma_v$ ones.  As we have seen with \PLI{}, $\sigma_v$ is a predictor of mass, albeit with significant scatter.

\begin{deluxetable*}{l l l l r r r r}
\tabletypesize{\scriptsize}

\tablecaption{Method Comparison \label{table:methodcomp}}
\tablewidth{0pt}
\tablehead{
\colhead{Case} & \colhead{Summary} & \colhead{Color} & \colhead{Catalog} & \colhead{$\mean{\epsilon}$\,\tablenotemark{1}} &
\colhead{$\epsilon \pm \Delta \epsilon$\,\tablenotemark{2}} & \colhead{$\Delta\epsilon$\,\tablenotemark{3}} & \colhead{$|\epsilon|-|\epsilon_{\mathrm{PL_C}}|$\,\tablenotemark{4}} 
}
\startdata
PLM{} 					& $M(\sigma_v)$ Power Law, \memb{}	& Red 		& Test 				&${0.128}$ & ${0.05 ^{+0.51 }_{-0.36}}$ & ${0.871}$ & ---  \\[1.5ex]
 					&							& 						& High-Mass Test 			&${0.093}$ & ${0.02 ^{+0.44 }_{-0.29}}$ & ${0.731}$ & ---  \\[1.5ex]
					
\PLI{}					& $M(\sigma_v)$ Power Law, \interloper{}		& Blue 		& Test 					&${0.508}$ & ${0.13 ^{+1.40 }_{-0.73}}$ & ${2.131}$ & ---  \\[1.5ex]
 					&							& 						& High-Mass Test 			&${0.409}$ & ${0.18 ^{+1.15 }_{-0.68}}$ & ${1.829}$ & ---  \\[1.5ex]

\MLv{}				& ML with $\vlos$				& Green 					& Test 					&${-0.052}$ & ${-0.12 ^{+0.40 }_{-0.27}}$ & ${0.670}$ &$-0.63$\\[1.5ex]
 					&							& 						& High-Mass Test 			&${-0.059}$ & ${-0.10 ^{+0.38 }_{-0.31}}$ & ${0.686}$ &$-0.47$\\[1.5ex]

\MLR{} 				& ML with R					& Orange 					& Test 					&${-0.016}$ & ${-0.08 ^{+0.39 }_{-0.28}}$ & ${0.670}$ &$-0.64$\\[1.5ex]
 					&							& 						& High-Mass Test 			&${-0.040}$ & ${-0.10 ^{+0.37 }_{-0.26}}$ & ${0.635}$ &$-0.49$\\[1.5ex]

\MLvR{}				& ML with $|\vlos|$ and $R$		& Brown					& Test 					&${0.078}$ & ${-0.04 ^{+0.56 }_{-0.34}}$ & ${0.899}$ &$-0.54$\\[1.5ex]
 					&							& 						& High-Mass Test 			&${-0.032}$ & ${-0.11 ^{+0.45 }_{-0.33}}$ & ${0.783}$ &$-0.42$\\[1.5ex]

\MLvsR{} 				& ML with $|\vlos|$, $\vsig$, \& $R$	& Purple 					& Test 					&${0.011}$ & ${-0.07 ^{+0.46 }_{-0.31}}$ & ${0.763}$ &$-0.61$\\[1.5ex]
 					&							& 						& High-Mass Test 			&${-0.044}$ & ${-0.09 ^{+0.36 }_{-0.29}}$ & ${0.649}$ &$-0.49$\\[1.5ex]

\enddata

\tablenotetext{1}{Mean fractional mass error.}
\tablenotetext{2}{Median fractional mass error $\pm$ 68\% scatter.}
\tablenotetext{3}{Width of $\epsilon$ 68\% scatter.}
\tablenotetext{4}{Mean difference between model and \PLI{} errors}
\end{deluxetable*}

By taking advantage of the full LOS velocity and projected radius distributions, the SDM approach to determining cluster mass from galaxy observables reduces the distribution of errors by roughly a factor of two, and also predicts masses well even in the cases where  \PLI{} catastrophically overpredicts, making it a valuable tool for probing cosmological models with observations of galaxy clusters.

\section{Discussion}
\label{sec:discussion}

 \begin{figure*}[]
  \centering
  \includegraphics[width=\textwidth]{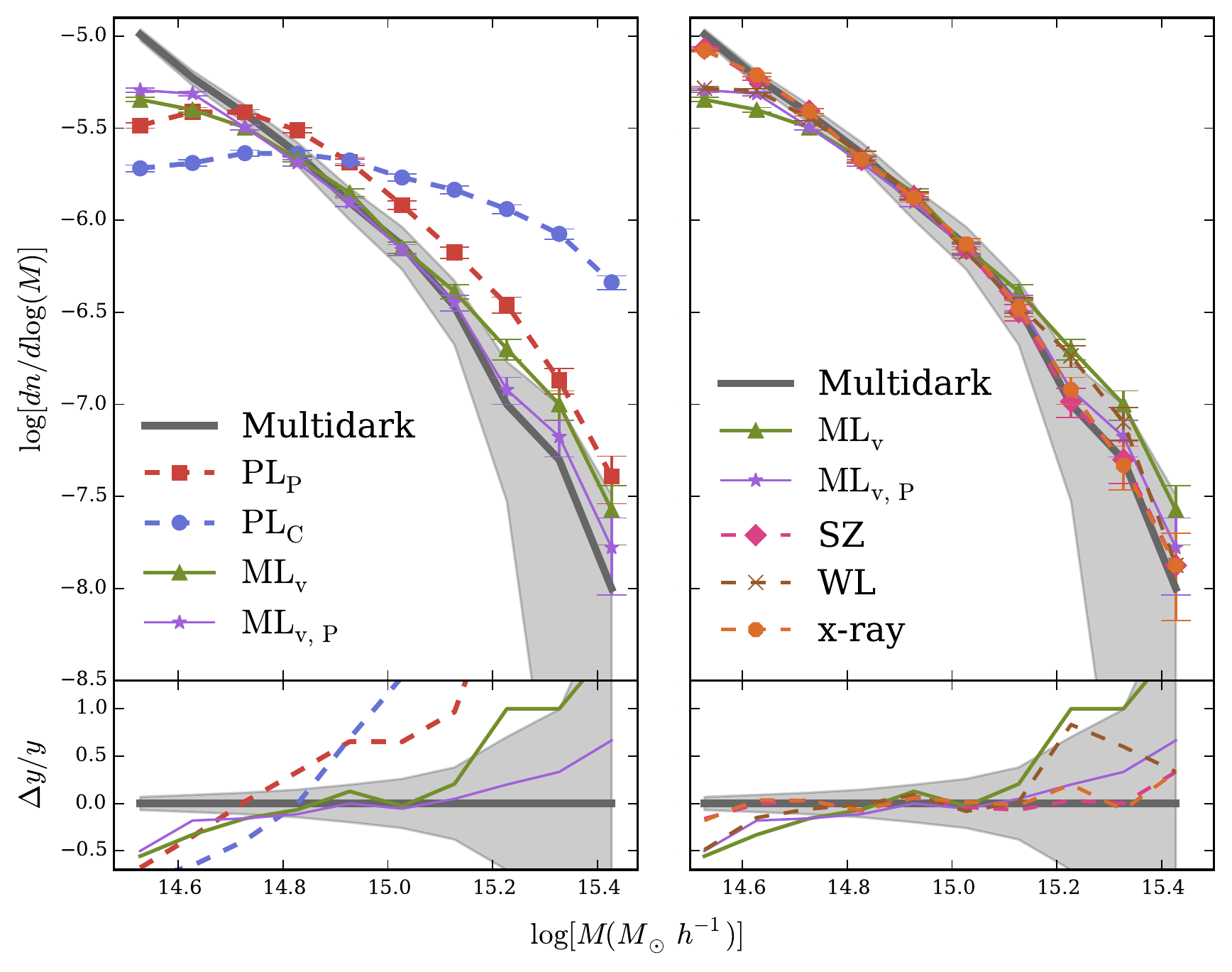}
  \caption{{Halo mass functions of dynamical cluster mass estimates with intrinsic scatter only (\memb{} Catalog) and intrinsic scatter plus observational selection effects (\interloper{} Catalog).  Any scatter in the mass-observable relationship, if uncorrected, will affect the observed halo mass function.   The large scatter associated with the power-law scaling relation (\PLM{}, red squares, and \PLI{}, blue circles) causes an upscatter at high masses, while ML methods (\MLvP{}, purple stars, and \MLv{}, green triangles) have a smaller intrinsic scatter and more accurately reproduce the true Multidark cluster abundance (dark gray solid curve).  While 6834 (7449) clusters contribute to the HMF for the \memb{} (\interloper{}) Catalog, a more moderate observation of 500 clusters yields larger Poisson error bars (light gray band).
  Right:  HMF of ML methods compared to mock HMF with the typical intrinsic scatter of Sunyaev-Zel'dovich (pink diamond), weak lensing (brown x), and x-ray (orange octagon) cluster masses.  The biases and the observational effects associated with SZ, WL, and x-ray masses may introduce additional scatter, causing the HMF to deviate further from the simulation HMF.  }}
  \label{fig:massfunction}
\end{figure*}

Reducing errors and eliminating biases in cluster mass measurements are crucial to utilizing clusters to discern and constrain cosmological models.  The halo mass function and its evolution are sensitive to cosmological parameters such as $\sigma_8$, $\Omega_M$, $\Omega_{\mathrm{DE}}$, and $w$ \citep[e.g.][]{2003A&A...398..867S, 2009ApJ...691.1307H, 2009ApJ...692.1060V, 2010ApJ...708..645R,2010MNRAS.406.1773M, 2011ARA&A..49..409A}. 
Therefore, accurate measurements of cluster abundance as a function of mass and redshift can be used to understand the underlying cosmology.  The limiting factor in constraining parameters and evaluating cosmological models with cluster counts, however, is in accurately connecting galaxy observables to halo mass to reproduce the halo mass function.  

Figure \ref{fig:massfunction} shows how the scatter and biases in each model affect the halo mass functions recovered by \PLM{}, \PLI{}, \MLv{}, {and \MLvP{} (SDM applied to the pure catalog with feature $|v_\mathrm{los}|$, as in \cite{Ntampaka2015})} in comparison to the simulation's true mass function.  The scatter about the scaling relation in \PLM{} coupled with the rapidly-declining shape of the mass function causes the abundant, low-mass clusters with high $\delta$ to populate the high-mass bins in the mass function, causing the upscattering at high masses.  This effect is exacerbated in \PLI{}, where the scatter about the scaling relation is much larger and the high-$\delta$ clusters may be catastrophically overpredicted (as shown in Figure \ref{fig:PDF}).  This {effect, known as Eddington bias \citep{1913MNRAS..73..359E},} alters the shape and amplitude of the measured halo mass function from the true value.  This results in \PLI{}'s measured mass function dramatically overreporting the number of high-mass clusters.  

{Any cosmological analysis of the HMF that employs such mass measurements must correct for this upscatter at high masses. Understanding the nature of the intrinsic scatter and observational selection effects is a crucial step to correct the observed HMF for Eddington bias.  Analytic approaches exist to correct for the simple case of lognormal scatter \citep[e.g.][]{2011PhRvD..83b3015M, 2014MNRAS.441.3562E}, while a more complicated scatter may be more difficult to correct.  Before correction for Eddington bias, the large scatter and errors associated with traditional power-law mass measurements lead to the failure to recover the true mass function, which limits the constraining power of dynamical mass measurements of galaxy clusters. } \PLM{}'s altered shape mimics the mass function of a simulation with a higher $\sigma_8$ and $\Omega_M$.  {This is particularly pronounced in the fractional difference, $\Delta y/y$, between the Multidark and mock HMFs, which shows that the presence of interlopers causes the PL HMF to deviate from the simulation HMF, particularly at high masses.}

{At the low mass end, the underabundance of clusters is not caused by Eddington bias, but is an artifact of the hard lower mass limits of the test catalogs.  This downscattering should not be interpreted as a dearth of low-mass clusters predicted by the PL and ML methods, but rather as a limitation of the test catalogs. }

{In addition to the halo mass functions from the methods highlighted in this work, mock HMFs that include scatter of other common cluster mass measurement techniques are included for comparison.   Cluster masses can be deduced from a variety of techniques, and here we show three different methods for determining cluster mass:  the Sunyaev-Zel'dovich (SZ) effect, weak gravitational lensing (WL), and x-ray.  
The SZ effect, first proposed by \cite{1972CoASP...4..173S} can be used to determine a temperature-weighted gas mass, and we model its intrinsic scatter according the \cite{2012ApJ...758...74B} scaling relation for $z=0$ with AGN feedback.
Weak gravitational lensing probes structure along the line-of-sight, and we model scatter in this technique according to the \cite{2011ApJ...740...25B} prescription for $z=0.25$, $M_{500c}\geq2.0\times10^{14}\,\Msolarh$ clusters.  X-ray observations can be used to infer a gas mass profile; scatter in this $M-Y_X$ relation of $\sigma_{\ln M}=0.06$ is adopted from \cite{2011MNRAS.416..801F}, and it should be noted that this is intrinsic scatter and does not include observational effects.  The mass-concentration relation from \cite{2013ApJ...766...32B} and the NFW density profile from \cite{1996ApJ...462..563N} are implemented to convert all masses to $M_{200c}$ for comparison. }

{Figure \ref{fig:massfunction} shows the  halo mass functions recovered by SZ, WL, and x-ray methods compared to the range of scatters achievable with SDM: \MLvP{} with a pure and complete cluster membership catalog and \MLv{} with a large cylindrical cut around each cluster allowing many interlopers.  It should be noted that the HMF presented assumes a complete large mock observation of 6834 (7449) clusters in the \memb{} (\interloper{}) Catalog.  Figure \ref{fig:massfunction} also shows the Poisson error associated with a more reasonable observation of 500 clusters.  Current cluster surveys \citep[e.g.][]{2016arXiv160306522D} contain on the order of hundreds of clusters, and the choice of $500$ clusters is chosen to show the errors accessible through current catalogs.  Note that the small number of high mass objects limit the accuracy with which the tail end of the HMF can be determined.  As is shown in, e.g., \cite{2016arXiv160201837N}, a binned HMF has the most power to resolve $\sigma_8$-$\Omega_m$ models at the lowest masses because, while high-mass clusters are sensitive to changes in these cosmological parameters, the Poisson error bars on these rare objects dominates.  For the mass ranges where the HMF can best resolve changes in $\sigma_8$ and $\Omega_m$, SDM produces a competitive HMF to these other mass proxies, though it has a larger deviation from the true HMF at the high mass tail.}

{However, it should be noted that these cluster mass methods utilize different wavelength observations with different systematic errors, biases, and limitations.  Therefore, while Figure \ref{fig:massfunction} shows that five different cluster mass techniques - PL, ML, SZ, x-ray, and WL - in a direct comparison, it should not be overly interpreted as a definitive guide to cluster mass measurement.  
For example, weak lensing is difficult and expensive to apply to high redshift clusters due to a lack of adequate background galaxies. 
Biases in x-ray and SZ cluster masses may arise because of nonthermal pressure support \citep[e.g.][]{1990ApJ...363..349E, 2004MNRAS.351..237R, 2009ApJ...705.1129L} (this bias is not modeled in Figure \ref{fig:massfunction} because this effect is typically corrected for, though uncertainty in the bias may produce further disagreement between observed and true HMF). 
When SZ masses are calibrated on simulation,  the calibration is dependent on correct modeling of the gas physics \citep[e.g.][]{2006ApJ...650..538N, 2012ApJ...758...74B}, which may also introduce a bias. }

{Dynamical and ML masses, however, can be directly compared as they are produced from the same data from the same mock catalog and are affected by the same observational selection effects.  From their direct comparison, it can be concluded that the ML method presented in this work is more competitive than a power-law scaling relation for decreasing errors in cluster mass measurements.}  While \MLv{} over predicts the abundance of high-mass clusters, the upscatter is smaller than \PLM{}'s.  \MLv{} provides a much better match to the simulation's true mass function across a larger mass range{, comparable to those of SZ, WL, and x-ray for the large mock observation of  $\approx (1 Gpc \,h^{-1})^3$.  This agreement with the true HMF} is primarily due to the small spread in errors associated with these methods; abundant, low-mass clusters tend not to be catastrophically overpredicted by {methods with small intrinsic scatter}.  The smaller errors produced in SDM's mass prediction results in a more accurate representation of the halo mass function, particularly at the high-mass end.  SDM's ability to more accurately recreate the true halo mass function makes it a valuable tool for producing cluster mass functions to evaluate cosmological models.  The predictive power of SDM to reproduce the true halo mass function and its implications for constraining cosmological parameters $\sigma_8$ and $\Omega_M$ will be explored in detail in an upcoming work.

{Section \ref{sec:appendix} explores how the aperture and, less directly, the purity and completeness of the cluster sample, affect the scatter in both power law and machine learning dynamical masses.  We find that the power law fit changes as a function of aperture, shallowing with smaller aperture.  When a large aperture is used, the distribution of errors at low masses is not lognormal, but is better described by a double Gaussian (see Figure \ref{fig:gauss}). }

{With the simple cylindrical cut and 2-$\sigma$ paring used in this work, mock cluster observations performed with a large aperture will tend to be more complete (compared to a mock observation made with a smaller aperture), with cluster members near the edges of the cluster being included in the sample.  Mock observations with a smaller aperture will tend to be more pure, with fewer interlopers contaminating the observation.  As we will show in Section \ref{sec:appendix}, SDM performs slightly better with a large aperture, showing a preference for completeness over one for purity.}

One may consider improving SDM mass predictions further by training and testing on features beyond simply $R$ and $\vlos$, applying a more accurate cluster interloper removal technique, or limiting the training sample to a particular subpopulation of galaxies.  Because elliptical galaxies preferentially reside in galaxy clusters \citep{1980ApJ...236..351D}, limiting the training sample to this population may provide a straightforward and natural approach to excluding many interlopers while still providing limited information about infalling matter.  But before such a training set can be explored and applied to observational data, there remains a need for a reliable training $N$-body simulation that is large, high resolution, and realistically populated with galaxies.

\section{Conclusions}
\label{sec:conclusions}

We compare cluster mass predictions from a standard $M(\sigma_v)$ power-law scaling relation to those generated by support distribution machines (SDMs), a machine learning class of algorithms that learn from a distribution of data to predict a scalar.

We focus on mass predictions for a mock catalog of impure and incomplete clusters.  This catalog is created from the publicly available Multidark MDPL1 simulation, with an intentionally-simplistic cylindrical cut imposed around the known centers of clusters with true mass $\geq 1\times 10^{14}\,\Msolarh$.  The aperture ($R_\mathrm{aperture}=1.6\,\Mpch$) and initial velocity cut ($v_\mathrm{cut}=2500\,\kms$) correspond to a typical radius and $2\times\sigma_v$ of a halo with mass $1\times 10^{15}\, \Msolarh$.  Velocity outliers beyond $2\sigma_v$ are iteratively pared until convergence, and only clusters with at least $20$ cluster members are kept in the sample.  This creates a catalog of clusters that are both impure (interlopers contaminate the clusters) as well as incomplete (some true cluster members are excluded from the sample).  A second catalog, both pure and complete, is also prepared for comparison.

Cluster masses are predicted in two ways:  in the PL approach, a standard $M(\sigma_v)$ power law is used to train and test, while in the ML approach, SDM is utilized.  Four feature sets are considered with SDM:  \MLv{} (absolute value of the line-of-sight velocity, $|\vlos|$), \MLR{} (galaxy projected distance from the cluster center, $R$), \MLvR{} ($|\vlos|$ and $R$), and \MLvsR{} ($|\vlos|$, $|\vlos|/\sigma_v$, and $R$).  Results for halos with true mass $M\geq3\times10^{14}\, \Msolarh$ are reported.  

Our main conclusions can be summarized as follows:
\begin{enumerate}
\item \MLv{} and \MLR{} (SDM with $|\vlos|$ feature only and SDM with $R$ feature only, respectively) reduce errors by 69\% compared to a power law applied to the same \interloper{} Catalog.  \item Further, {though a simple cylindrical cut causes significant scatter in the $M(\sigma_v)$ power law compared to when the cluster membership is perfectly known, both SDM} methods each outperform \PLM{}, a power law applied to a catalog with pure and complete clusters.  Compared to this ideal power law, \MLv{} and \MLR{} each reduce error by 23\%. 
\item Though it reduces error width, \MLvR{} (SDM with $|\vlos|$ and $R$) systematically underpredicts the highest-mass clusters.  It is identified as a disfavored method.
\item \MLvsR{} (SDM with $|\vlos|$, $|\vlos|/\sigma_v$, and $R$) minimizes the bias for the high-mass clusters ($M\geq7\times10^{14}\, \Msolarh$).  It reduces error by 64\% and 12\% compared to \PLI{} and \PLM{}, respectively.
\item In some instances, a higher-than-expected $\sigma_v$ causes a catastrophic overprediction by method \PLI.  The ML methods, however, predict reasonable masses for even these outliers.
\end{enumerate}

The SDM approach to determining cluster mass from galaxy observables reduces errors by more than a factor of two {compared to a standard power-law scaling approach applied to a cluster catalog with impure, incomplete cluster membership information.  SDM predicts cluster} masses well even when a traditional $M(\sigma_v)$ approach fails.  Additionally, this technique works well even with catalogs of impure and incomplete clusters created with a simplistic cylindrical cut about the cluster center.  Ultimately, high-resolution, large-volume simulations are needed for training before SDM can be applied to observation.   With such a simulation for training, the reduced errors and more accurate predictions for impure, incomplete, nonvirialized systems makes SDM a valuable tool for constraining cosmological models.

\acknowledgments{
We thank Nicholas Battaglia, Arthur Kosowsky, Rachel Mandelbaum, and Crist\'{o}bal Sif\'{o}n for their valuable feedback on this manuscript.
This work is supported in part by DOE DE-SC0011114 grant. 
The CosmoSim database used in this paper is a service by the Leibniz-Institute for Astrophysics Potsdam (AIP).
The MultiDark database was developed in cooperation with the Spanish MultiDark Consolider Project CSD2009-00064.
The Bolshoi and MultiDark simulations have been performed within the Bolshoi project of the University of California High-Performance AstroComputing Center (UC-HiPACC) and were run at the NASA Ames Research Center. The MultiDark-Planck (MDPL) and the BigMD simulation suite have been performed in the Supermuc supercomputer at LRZ using time granted by PRACE.  }

\appendix 
\label{sec:appendix}

Here, we explore how our choices of $R_\mathrm{aperture}$ and $v_\mathrm{cut}$ affect the PL and ML predictions and results.  Two new catalogs are prepared to correspond to a $3\times10^{14}\,\Msolarh$ cluster ($R_\mathrm{aperture}=1.1\,\Mpch$ and $v_\mathrm{cut}=1570\,\kms$, denoted ``Small Aperture'') and $3\times10^{15}\,\Msolarh$ cluster ($R_\mathrm{aperture}=2.3\,\Mpch$ and $v_\mathrm{cut}=3785\,\kms$, denoted ``Large Aperture'').  The \interloper{} Catalog used in the main body of this work has been renamed ``Medium Aperture'' for clarity.  As before, a 2-$\sigma$ iterative paring scheme is applied to the initial cylindrical cut.  With the exception of the $R_\mathrm{aperture}$ and $v_\mathrm{cut}$ values, the methods described in Sec.~\ref{sec:methods} are followed.  These catalogs, along with the \memb{} Catalog, are summarized in Table \ref{table:appendixCat}.

\begin{table}[t]
\begin{center}
\caption{{Catalog Summary} \label{table:appendixCat}}
\begin{tabular}{l l r r r r r r} 
\tableline
\tableline
\multicolumn{1}{l}{Catalog} 	&\multicolumn{1}{l}{Type}		& \multicolumn{1}{l}{$R_\mathrm{aperture}$} 		& \multicolumn{1}{l}{\vcut}	&  \multicolumn{1}{l}{$\sigma_{15}$} 	&  \multicolumn{1}{l}{$\alpha$} 	\\

\multicolumn{1}{l}{Name} 				&\multicolumn{1}{l}{}  		& \multicolumn{1}{l}{$(\Mpch)$} 	& \multicolumn{1}{l}{$(\kms)$}&  \multicolumn{1}{c}{$(\kms)$} 				&  \multicolumn{1}{c}{} \\
\tableline\\[1ex]

Small 					& PL Train 					& 1.1 							&1570		& 569							& 0.209	\\
Aperture & & & & & \\[1.5ex]
Medium				& PL Train 					& 1.6 							&2500		& 895							& 0.384	\\
Aperture & & & & & \\[1.5ex]
Large 					& PL Train 					& 2.3								&3785		& 900							& 0.400	\\
Aperture & & & & & \\[1.5ex]
\memb{}				& Train						& ---								& ---			& 1244							& 0.382 \\[1.5ex]

\tableline
\end{tabular}

\end{center}
\end{table}

\begin{figure*}[]
\begin{center}
\begin{tabular}{@{}c@{}c@{}c@{}}
     \includegraphics[height = 0.28 \textwidth]{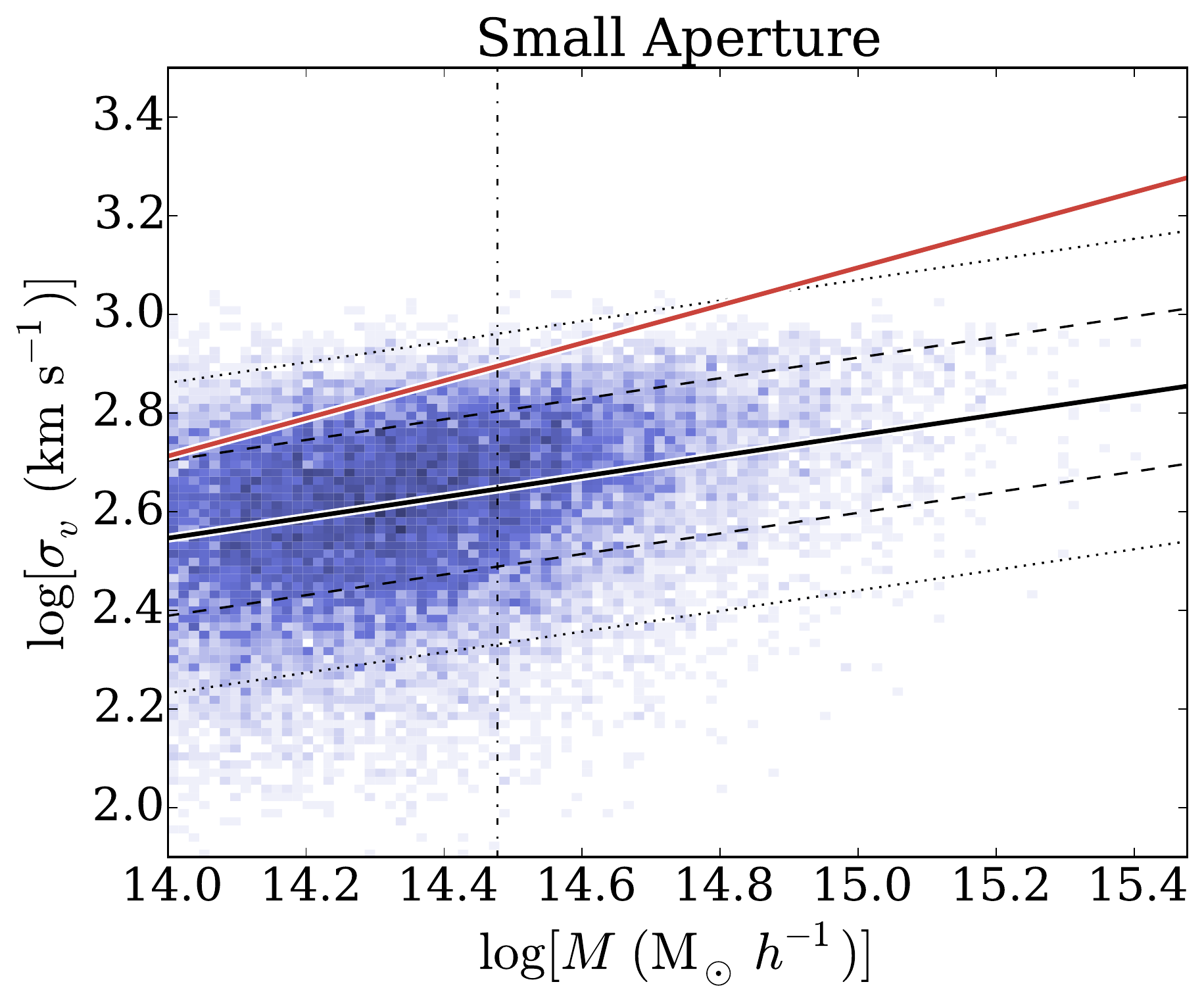}  & \includegraphics[height = 0.28 \textwidth]{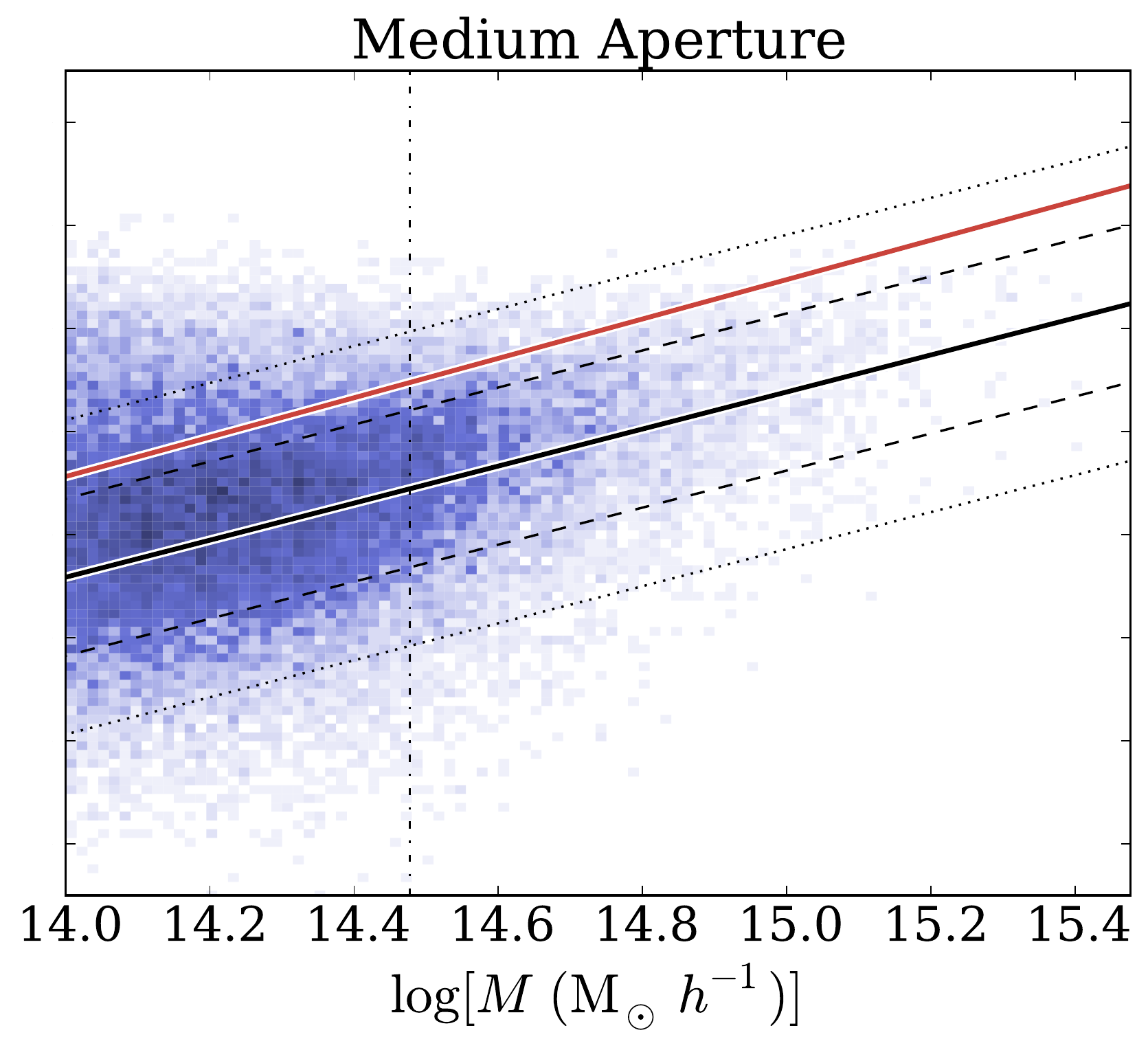} &  \includegraphics[height = 0.28 \textwidth]{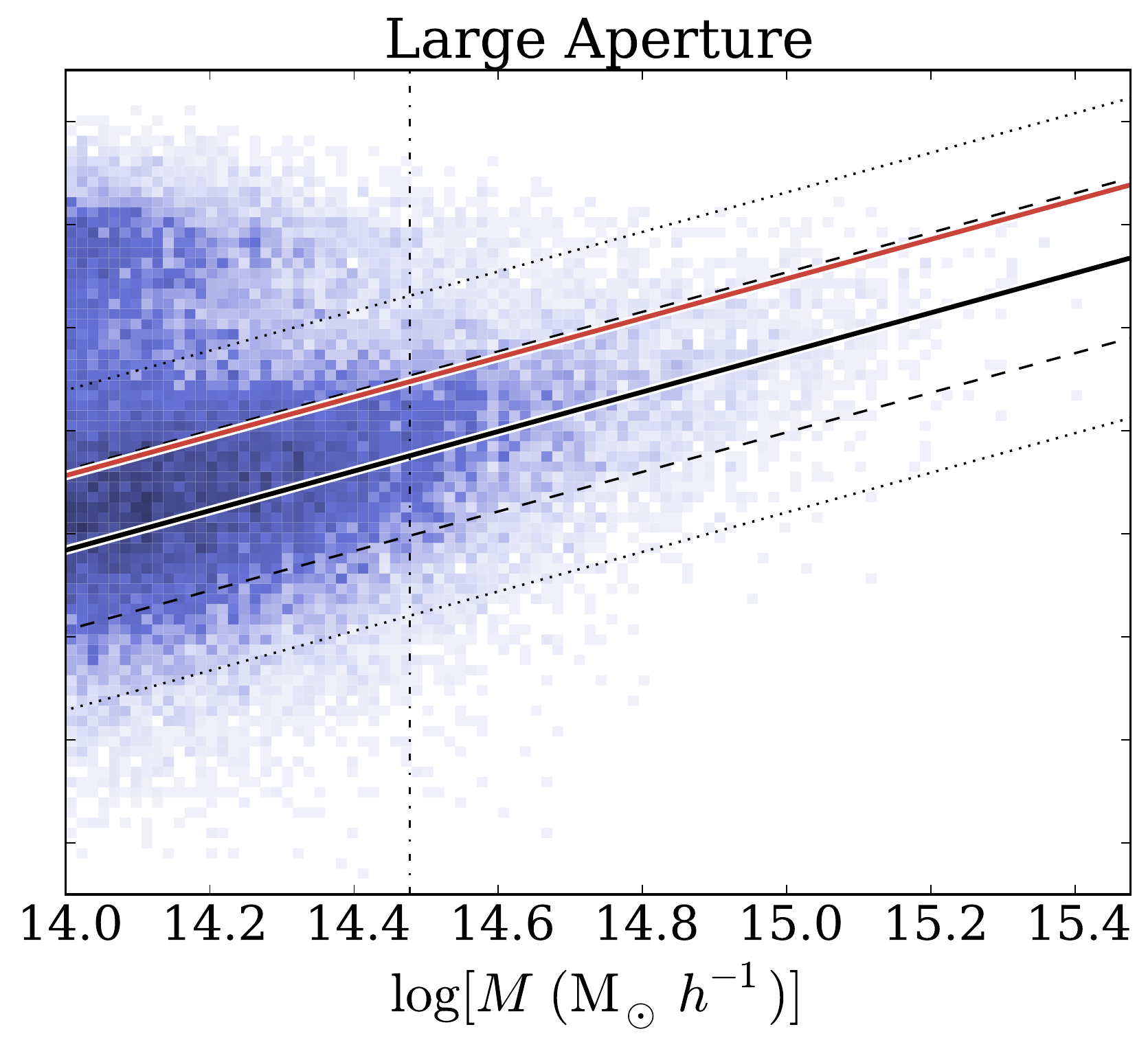}  \\   \end{tabular}
       \caption{Left: Small Aperture Catalog's LOS velocity dispersion of galaxies, $\sigma_v$, vs.~cluster mass, $M$, shown as a 2D histogram.  Only clusters above $3\times10^{14}\, \Msolarh$ (black dash dotted) are used to determine the best fit power law (black solid); the small aperture and $v_\mathrm{cut}$ lead to smaller-than-expected $\sigma_v$'s for the high mass halos and result in a shallow fit.  The $M(\sigma_v)$ fit for pure and complete clusters (\PLM{}, red) is overlaid for reference.  Center:  Medium Aperture Catalog.  If the lognormal scatter in $\sigma_v$ was consistent across the entire mass range, the 1- and 2-$\sigma$ errors (black dashed and dotted, respectively) calculated at the high-mass end would describe the scatter in $\sigma_v$ even at low masses.  However, a clear trend emerges, with increased scatter in $\sigma_v$ at lower masses.  Right:  Large Aperture Catalog.  The slope of the power law has steepened.  This is due to the larger $R_\mathrm{aperture}$ and $v_\mathrm{cut}$ used for this catalog, which capture more true members of the high-mass clusters, allowing these objects to be more accurately described.  Though the high-mass clusters are now well-represented by their measured $\sigma_v$, a clear second population emerges at low mass and high $\sigma_v$, with $20\%$ of halos with $M<3\times10^{14}\,\Msolarh$ lying above the 2-$\sigma$ dotted line. }
	\label{fig:msigA}
      \end{center}
\end{figure*}

\begin{figure*}[!tbh]
\begin{center}
\begin{tabular}{c c}
     
     	\includegraphics[width=0.5\textwidth]{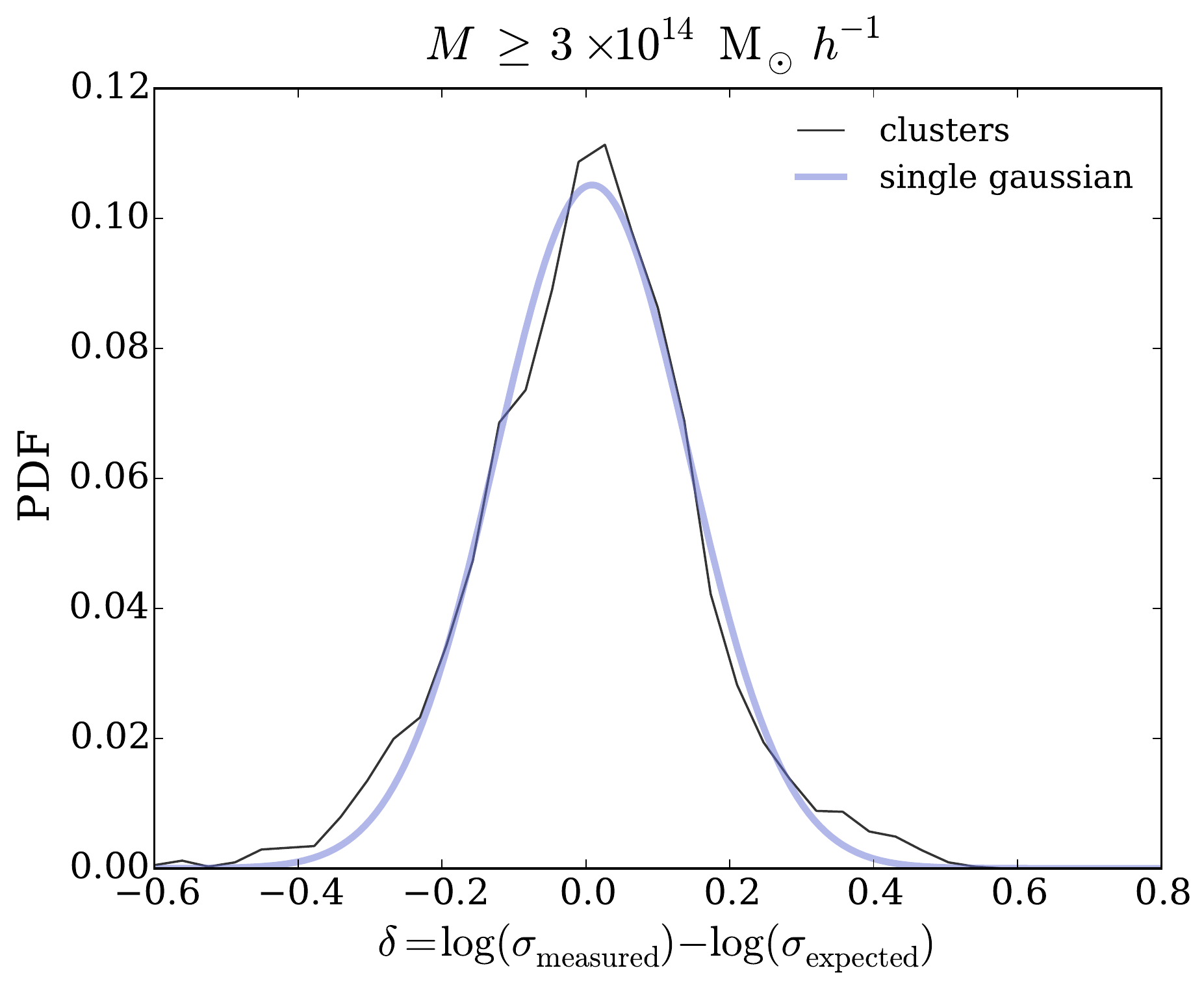} &\includegraphics[width=0.5\textwidth]{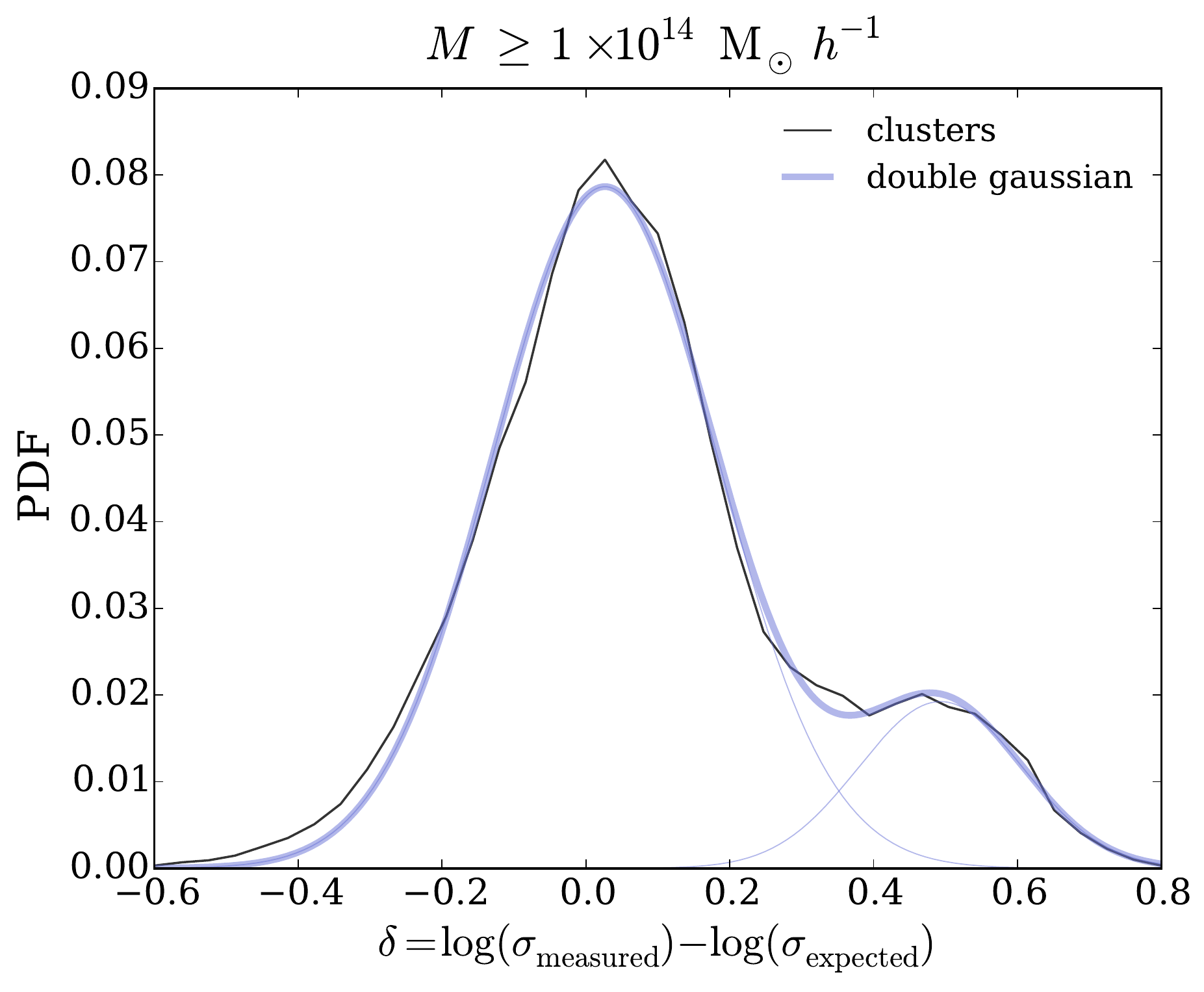} \\
         
  \end{tabular}
       \caption{ Left:   PDF of residual, $\delta$, for the Large Aperture Catalog.  With a lower mass cut of $M=3\times10^{14}\, \Msolarh$, the PDF of clusters' $\delta$ (thin black) is well-described by a single Gaussian (thick blue).  Right:  When the mass limit of the Large Aperture Catalog is lowered to $M=1\times10^{14}\, \Msolarh$, the PDF is better-described by a double Gaussian.  Observational methods for identifying members of this outlier population will be explored in a later work.}
      \label{fig:gauss}
      \end{center}
\end{figure*}

Figure \ref{fig:msigA} shows how the choice of $R_\mathrm{aperture}$ and $v_\mathrm{cut}$ affect the power-law fits.  This two-dimensional histogram of $\sigma_v$ vs.~$M$ shows that the best fit $\alpha$ and $\sigma_v$, as well as the scatter about the best fit line, changes as a function of initial cylinder size.  Overlaid on the two-dimensional histogram is a best fit with 1- and 2-$\sigma$ lognormal errors, calculated for clusters with mass above $3\times10^{14}\, \Msolarh$ and extrapolated down to lower masses.  Additionally overlaid is the best fit power law for \PLM{}.

When the Small Aperture cuts are applied, this overly-small cylinder clips the $\sigma_v$ values at the high mass.  This leads a shallow slope ($\alpha=0.209$) and small velocity dispersion associated with a $10^{15}\,\Msolarh$ cluster ($\sigma_{15}=569\,\kms$).  In contrast, a large cylindrical fit increases scatter at the low-mass end.  The resulting fit for the Large Aperture Catalog is steep ($\alpha=0.384$) and has a higher normalization ($\sigma_{15}=895\,\kms$) caused by the many high-$\sigma_v$ objects and the substantial fraction of outliers above the 2-$\sigma$ line.  These catalogs and fits are summarized in Table \ref{table:appendixCat} for reference.

\begin{figure*}[!tbh]
\begin{center}
\begin{tabular}{c c c}
     
     	\includegraphics[width=0.3\textwidth]{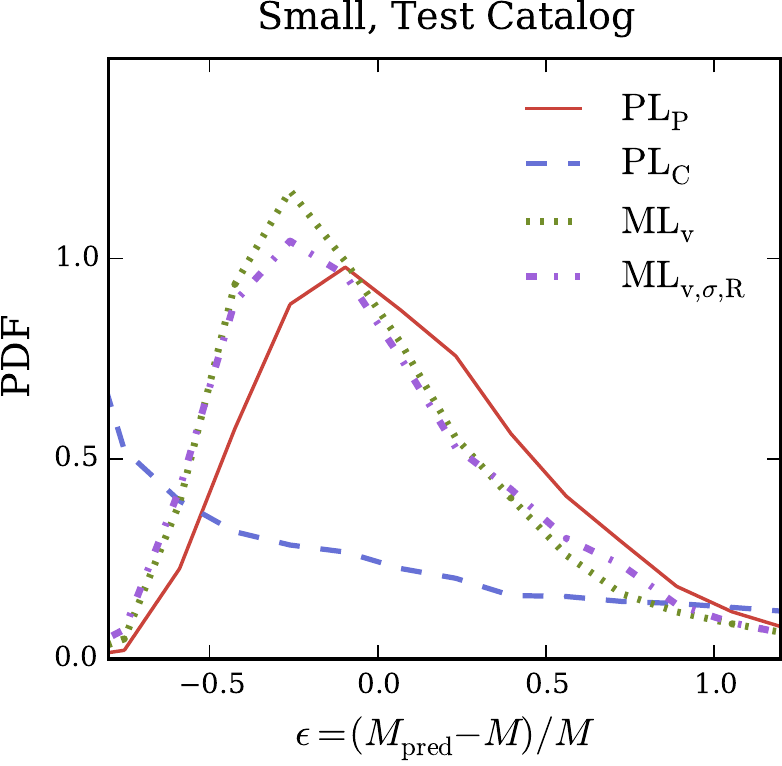} &\includegraphics[width=0.3\textwidth]{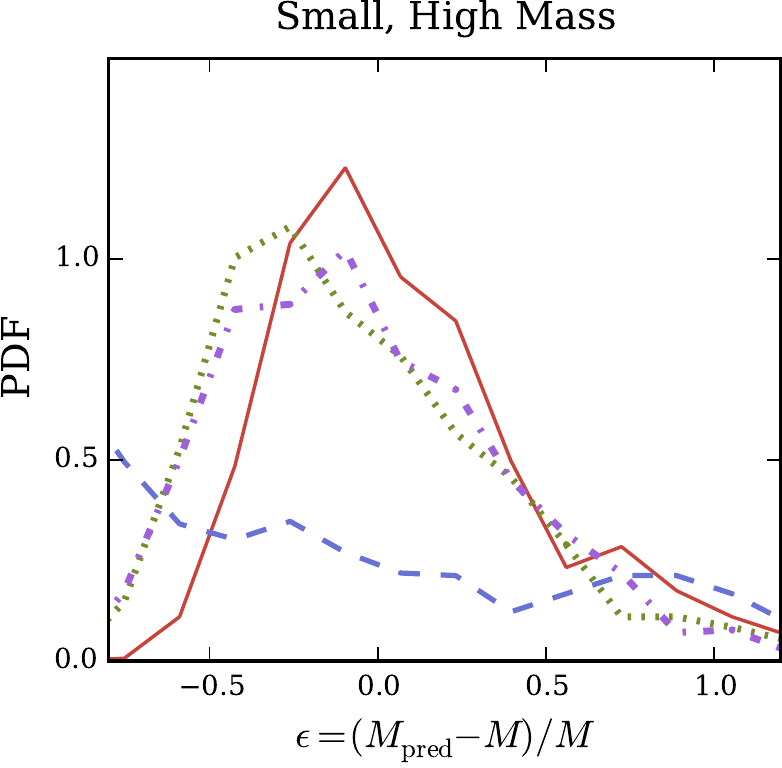}&\includegraphics[width=0.3\textwidth]{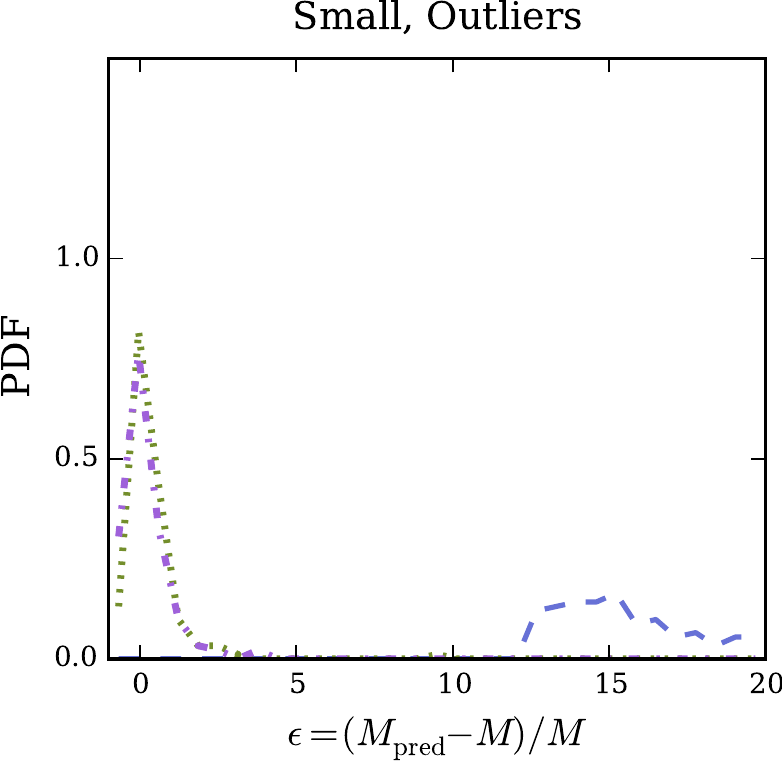}  \\
	\\[1.5ex]
         \includegraphics[width=0.3\textwidth]{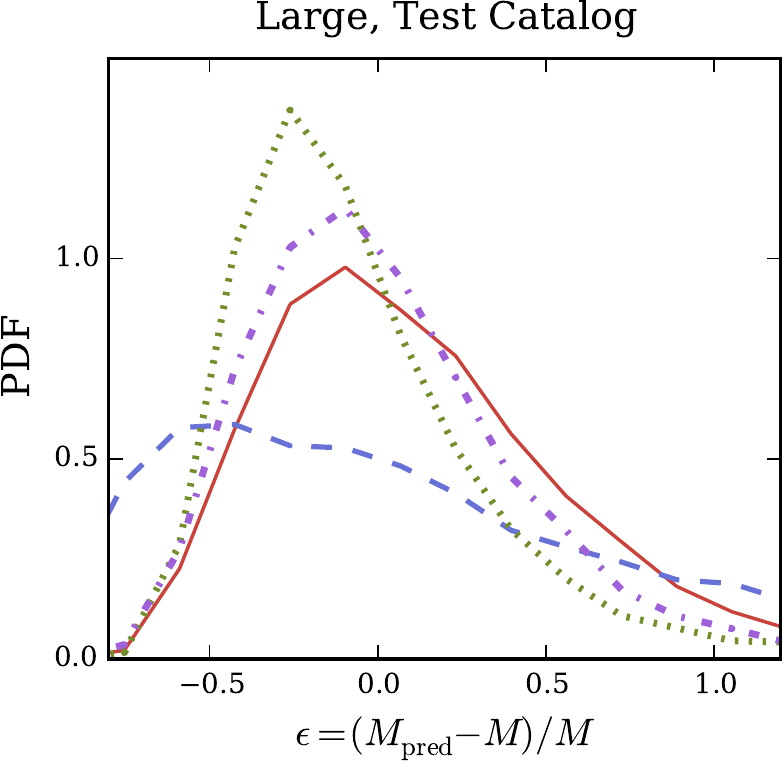} &\includegraphics[width=0.3\textwidth]{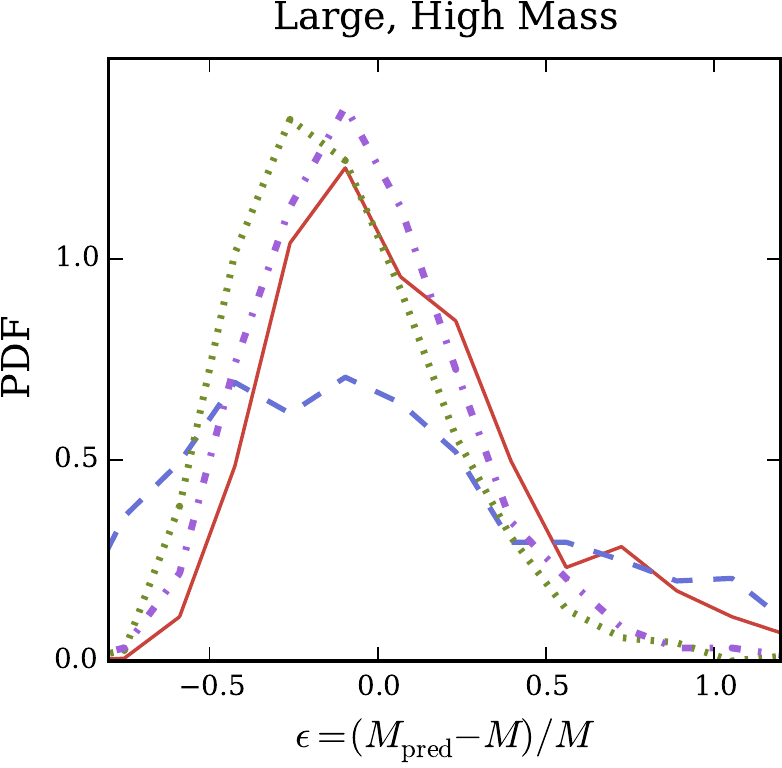} &\includegraphics[width=0.3\textwidth]{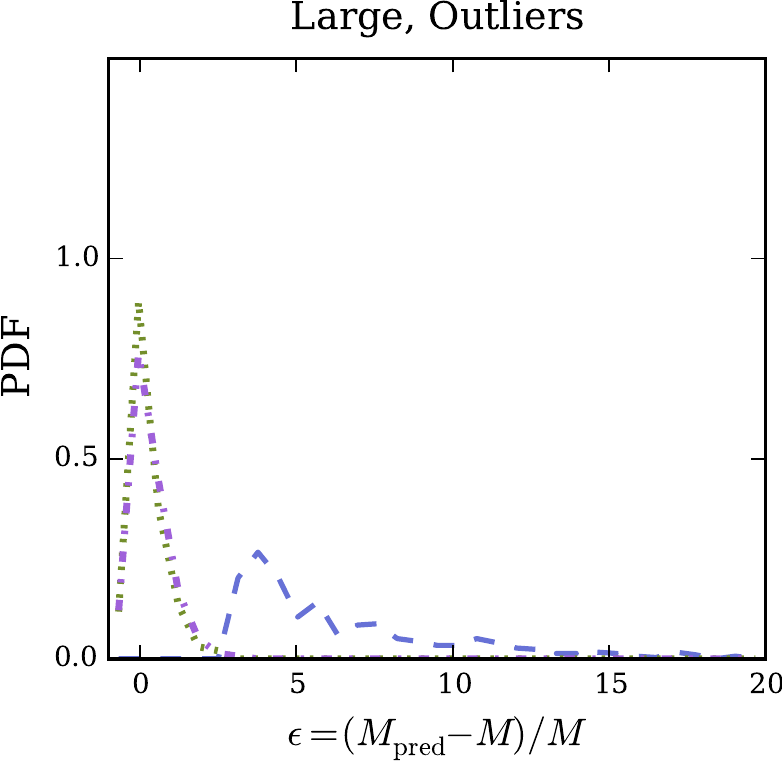} \\
         
  \end{tabular}
       \caption{Top Left:  PDF of errors for the Small Aperture Catalog.   When this small cut is imposed on the mock observation, the shallow slope of the fit causes large-negative-$\delta$ population to be underpredicted in mass by an order of magnitude or more, creating the abundance of clusters with $\epsilon \lesssim 0.1$.  Top Center:  Small Aperture Catalog, high mass halos only ($M\geq7\times10^{14}\,\Msolarh$) has a similar abundance of underpredicted halo masses.  Top Right:  PDF of errors for the high-error objects.  The shallow Small Aperture fit also results in a number of catastrophically overpredicted clusters.  SDM, however, predicts reasonable masses for even these outliers.  Bottom Left: PDF of errors for the Large Aperture Catalog.  The large cut leads to more interlopers, but SDM predicts better than a scaling relation applied to a pure and complete catalog.   Bottom Center:  Large Aperture Catalog, high mass halos only.  Bottom Right:  PDF of high-error objects for the Large Aperture Catalog.  SDM predicts reasonably accurate masses here, though a power-law scaling relation fails catastrophically.}
      \label{fig:PDFAppendix}
      \end{center}
\end{figure*}

As the Large Aperture Catalog's cuts are used to probe lower masses, a bimodal distribution emerges with a second population of clusters residing far above the best fit; this second population is visible in Figure \ref{fig:msig}.  These high-$\sigma_v$, low-mass objects increase scatter at the low-mass end.  More worrisome, they have a velocity dispersion typically associated with clusters of roughly an order of magnitude larger in mass.  Of halos with $M\geq3\times10^{14}\,\Msolarh$, 3\% reside above the $+2\sigma$ dotted line and 3\% reside below the $-2\sigma$ dotted line.  However, of halos with $1\times10^{14}\, \Msolarh \leq M < 3\times10^{14}\,\Msolarh$, 20\% reside above $+2\sigma$ and 3\% below $-2\sigma$.  The best fit and lognormal scatter found for the higher-mass clusters in the Large Aperture Catalog is clearly not descriptive of the lower-mass clusters.

To further explore this outlier population, we will consider the residual, $\delta$ (Equation \ref{eq:residual}).  Figure \ref{fig:gauss} shows that the Large Aperture Catalog has a residual PDF is adequately described by a single Gaussian, parameterized by
\begin{equation}
	\mathrm{PDF} \propto \exp\left[{\frac{-(\delta-\mu)^2}{2 \, \sigma_\mathrm{gauss}^2}}\right],
	\label{eq:gauss}
\end{equation}
with best fit width $\sigma_\mathrm{gauss}=0.13$ and a nearly-zero offset, $\mu=0.01$.  

However, when the lower mass limit of this Large Aperture Catalog is decreased to $1\times10^{14}\, \Msolarh$, the  $\delta$ PDF is better described by the sum of two Gaussians, as shown Figure \ref{fig:gauss}.  The relative amplitude and width of high-$\delta$ Gaussian is dependent on the minimum mass cut applied to the catalog, and our choice of $1\times10^{14}\, \Msolarh$ is for illustrative purposes only.  Note, however, that the zero-centered Gaussian has $\sigma_\mathrm{gauss} = 0.16$ and $\mu=0.03$, comparable to the single Gaussian fit found previously.  This is suggestive that a single lognormal scatter describes the population that is well-characterized by the $M(\sigma_v)$ power law, while a second population of high-$\sigma_v$ outliers emerges at low masses.  Exploring observational methods for describing and identifying members of this outlier population will be considered in future work.

Figure \ref{fig:PDFAppendix} shows that the resulting large scatter produces \PLI{} error PDF that is wide and flat as before, with the shape of the \PLI{} PDF dependent on the cylindrical cut parameters.  For the Small Aperture Catalog, the shallow fit coupled with the large number of clusters with large negative $\delta$ contribute to the substantial population of clusters being underestimated by an order of magnitude or more ($\epsilon\lesssim-0.1$).  SDM produces a slightly wider error distribution for this small initial cylinder compared to the Medium Aperture cuts, though still reducing $\Delta\epsilon$ compared to both \PLI{} and \PLM{}.  Distributions of error as a function of mass are comparable to those seen in Figure \ref{fig:MLsummary}, regardless of the training catalog, though $\mean{\epsilon}$ tends to decrease and $\Delta\epsilon$ tends to widen for small initial cylinders.

As before, there are also a number of catastrophically overpredicted clusters by applying the \PLI{} scaling relation to the Small Aperture Catalog.  These overpredicted objects are identified by their residual relative to the lognormal scatter: $\delta \geq 1.5\times\sigma_\mathrm{gauss}$.  The shallow slope leads to the overprediction being much more pronounced.  However, Figure \ref{fig:PDFAppendix} shows that, even in this case, SDM predicts reasonably accurate masses for these objects. 

The population of high-$\sigma_v$, low-mass, high-$\delta$ objects in the Large Aperture Catalog similarly produces a substantial number of catastrophically overpredicted clusters.  These large-$\epsilon$ objects shown in Figure \ref{fig:PDFAppendix} are also well-predicted by SDM.  While the \PLI{} gives a large range of errors, SDM can more accurately predict these cluster masses despite overly-large or small cylindrical cuts that contribute to significant impurity or incompleteness in the mock clusters.

\MLv{} and \MLvsR{} produce the smallest $\Delta\epsilon$ when the initial cylinders are large, with $\Delta\epsilon=0.670$ and $0.763$, respectively, for the Medium Aperture Catalog and $\Delta\epsilon=0.660$ and $0.752$ for the Large Aperture Catalog.  The Small Aperture Catalog error distribution is wider: $\Delta\epsilon=0.809$ and $0.898$.  However, in all cases except \MLvsR{} applied to the Small Aperture cylinder, the width of error distribution is narrower than the \memb{} Catalog power law, which has $\Delta\epsilon=0.871$.  SDM performs better with impurity over incompleteness, with larger cylinders producing slightly more accurate mass predictions.

Errors produced by a power-law scaling relation are clearly dependent on the choices of $R_\mathrm{aperture}$ and $v_\mathrm{cut}$, sometimes catastrophically overpredicting cluster masses.  Though a standard power-law scaling fits and error distributions are sensitive to choices in cuts, SDM can predict accurately under a wide range of scenarios, provided the training and test data have the same imposed cuts.

\end{document}